\documentstyle[12pt]{JHEP3}
\newcommand{\bej}[1]{ \begin{equation}\label{#1} }
\newcommand{\eej}{\end{equation}}
\newcommand{\beaj}[1]{\begin{eqnarray}\label{#1} }
\newcommand{\eeaj}{\end{eqnarray}}
\newcommand{\equ}[1]{(\ref{#1})}
\def\ZZZ{{\hskip-3pt\hbox{ Z\kern-1.6mm Z}}}
\def\zzz{{\hskip-3pt\hbox{ z\kern-1mm z}}}

\def\a{\alpha}

\def\b{\beta}

\def\cA{{\cal{A}}}
\def\cL{{\cal{L}}}
\def\cP{{\cal{P}}}

\def\d{\delta}

\def\f{\frac}

\def\hs{\hspace}

\def\lf{\left}
\def\l{\lambda}

\def\lan{\langle}

\def\m{\mu}
\def\n{\nu}
\def\na{\nabla}
\def\nn{\nonumber}
\def\o{\omega}
\def\O{\Omega}
\def\p{\phi}

\def\pa{\partial}
\def\pr{\prime}

\def\ra{\rightarrow}
\def\ran{\rangle}

\def\ri{\right}

\def\tr{\textrm}

\def\vk{\vec{k}}
\def\vp{\varphi}

\newcommand{\be}{\begin{equation}}
\newcommand{\ee}{\end{equation}}
\newcommand{\bea}{\begin{eqnarray}}
\newcommand{\eea}{\end{eqnarray}}

\title{
Hydrodynamics from the D1-brane}
\author{
Justin R. David\footnote{On lien from Harish-Chandra Research Institute, Allahabad.} $^{a}$,
Manavendra Mahato$^{b}$ and Spenta R. Wadia $^{b, c}$ \\
$^a$Centre for High Energy Physics,
Indian Institute of Science, \\
Bangalore 560012, India.\\
$^b$ Department of Theoretical Physics, 
 Tata Institute of Fundamental Research, \\
 Homi Bhabha Road, Mumbai 40005, India.\\
$^c$ International Centre for Theoretical Sciences, TIFR,\\
 Homi Bhabha Road, Mumbai 40005, India.\\
\email{justin@cts.iisc.ernet.in, manav@theory.tifr.res.in, wadia@theory.tifr.res.in}
}

\abstract{
We study the hydrodynamic properties of
strongly coupled $SU(N)$ Yang-Mills theory of the   D1-brane at finite temperature
in the framework of gauge/gravity duality.
The only non-trivial viscous  transport coefficient in
$1+1$ dimensions is the bulk viscosity.
 We evaluate the bulk viscosity
by isolating the  quasi-normal mode corresponding to the sound channel for the gravitational  background of
 the D1-brane. We find that the ratio of the bulk viscosity to the entropy density to be
 $1/4\pi$. This ratio continues to be  $1/4\pi$ also  in the regime when the D1-brane
Yang-Mills theory is
dual to the gravitational background of the fundamental string.
Our analysis shows that this ratio is 
equal to $1/4\pi$ for a class of gravitational backgrounds dual to field  theories in
$1+1$ dimensions obtained by considering D1-branes at cones over Sasaki-Einstein 7-manifolds.
}
\preprint{TIFR/TH/09-01}

\begin{document}

\section{Introduction}

Recent studies of  gauge theories using the framework of the AdS/CFT correspondence has
revealed that gauge theories at large 't Hooft coupling and at long-distances and low-frequencies
can be described  by fluid mechanics 
\cite{Policastro:2001yc,Policastro:2002se,Policastro:2002tn,Herzog:2002fn,
Kovtun:2003wp,Herzog:2003ke,Kovtun:2004de}
\footnote{For a review and a complete list of references 
please see \cite{Son:2007vk}}.
A hydrodynamic description implies that the correlation functions of components of
stress-energy tensor or conserved currents are fixed once a few transport coefficients are known.
These transport coefficients have been evaluated for several examples of
gauge theories at strong coupling using the AdS/CFT correspondence.
Recently, transport coefficients for non-linear hydrodynamics have been
obtained \cite{Bhattacharyya:2008jc,Baier:2007ix}.
These studies indicate that for field theories which admit a gravity dual, the
ratio of shear viscosity $\eta$  to the entropy density  $s$ at strong t'Hooft coupling is universal \cite{Kovtun:2003wp,Buchel:2003tz} and is given by 
\begin{equation}
 \label{shear-entropy}
\frac{\eta}{s} = \frac{1}{4\pi}.
\end{equation}
Most studies of this ratio have been focused on asymptotic anti-de Sitter 
backgrounds in various dimensions.
The gauge/gravity correspondence also applies to stacks of D$p$-branes for arbitrary $p$ \cite{Boonstra:1998mp,Itzhaki:1998dd}\footnote{For a review and recent developments
on holography for the non-conformal case, please see \cite{Kanitscheider:2008kd}.}.
It was shown in \cite{Kovtun:2003wp,Mas:2007ng} that the ratio of shear viscosity to entropy density
for  theories on Dp-branes
for $p\geq 2$ also is $1/4\pi$ \footnote{Other non-conformal systems were studied in 
\cite{Benincasa:2005iv,Benincasa:2006ei,Buchel:2007mf}}.
An investigation of transport coefficients for  the theory on the D1-brane is missing
in the literature.

In this paper, we begin a study of the hydrodynamic behaviour of field theories
in $1+1$ dimensions which
admit a gravity dual.
What makes one spatial dimension special is the absence of shear.
In fact, for the conformal field theories in
$1+1$ dimensions, there are no transport coefficients. This is because there are no
non-trivial components of the symmetric traceless stress-energy tensor in these dimensions.
The stress tensor is that of a perfect fluid. Thus, to study non-trivial transport properties of field theories in
$1+1$ dimensions, it is necessary to study the non-conformal case.
The only viscous transport coefficient for  non-conformal field theories in $1+1$ dimensions is the
bulk viscosity.
The simplest example of such a non-conformal field theory is the theory on the
D1-brane, the $1+1$ dimensional $SU(N)$ gauge theory with $16$ supersymmetries.
It can be obtained as a dimensional reduction of ${\cal N}=4$ SYM from $3+1$ dimensions.
We consider this theory at finite temperature.
It admits  dual gravity descriptions in two regimes
\begin{eqnarray}
 \label{regimes}
&  & (i) \quad \sqrt{\lambda} N^{-2/3} << T << \sqrt{\lambda}, \qquad \hbox{and}  \\ \nonumber
& & (ii)\quad \sqrt{\lambda}{N^{-1}} << T<< \sqrt{\lambda} N^{-2/3}.
\end{eqnarray}
Here,  the 't Hooft coupling is denoted by $\lambda = g_{YM}^2N$ and $T$ is the temperature.
In regime $(i)$, the gravity dual is the background of that of the non-extremal D1-brane, while in
regime $(ii)$, the dual is that of non-extremal fundamental string.

One  reason why  transport coefficients were not studied for backgrounds corresponding to the D1-brane
is that  the gauge invariant fluctuations of supergravity fields  found for the case of D$p$ brane with
$p\geq2$ cannot be extended to $p=1$.
In this paper, we isolate the gauge invariant fluctuation
of the graviton and the dilaton which corresponds to the sound channel.
From its equation of motion,
we find the following dispersion
relation for its quasinormal mode
\begin{equation}
 \label{dispersion}
\omega= \frac{q}{\sqrt{2}} - \frac{i}{8\pi T} q^2.
\end{equation}
This dispersion relation for the quasinormal mode
 remains the same for both the D1-brane background as well as for the F1-string
background.
We show that the retarded two point functions  of components of stress tensor has a pole at the value of
$\omega$ corresponding to the sound mode.
From the universal properties of hydrodynamics in $1+1$ dimension, we find that the  pole is
given by
\begin{equation}
 \label{gen-pole-structure}
\omega = v_s q - i \frac{\xi}{2(\epsilon +P) } q^2.
\end{equation}
Here, $v_s$ is the speed of sound in the medium, $\epsilon$ the energy density and $P$ the pressure.
By comparing the dispersion relation (\ref{dispersion}) and the above equation, we can read out the following
properties of the D1-brane gauge theory in both the regimes given  in (\ref{regimes}).
\begin{equation}
 \label{result}
v_s = \frac{1}{\sqrt{2}}, \qquad \frac{\xi}{s} = \frac{1}{4\pi},
\end{equation}
where $\xi$ is the bulk viscosity and $s$ is the entropy density. 
As a cross check of our calculations we  use the Kubo's formula to evaluate the bluk viscosity and show that
the ratio  $\xi/s$  is given by $1/4\pi$.

It is curious that the ratio of the bulk viscosity to entropy density has the same value as that of
ratio $\eta/s$ for strongly coupled field theories which admit gravity duals in higher dimensions.
Also the fact that the ratio $\xi/s$ remains $1/4\pi$ for both the D1-brane and the F1-string suggests
that this ratio might be universal for a class of gravitational backgrounds.
We show that the this ratio continues to be $1/4\pi$ for the class of $1+1$ dimensional 
 non-conformal field theories on D1-branes at cones over Sasaki-Einstein 7-manifolds.

The organization of the paper is as follows: In the next section, we briefly review the gauge/gravity
correspondence and thermodynamics for the case of the D1-brane. In section 3, we discuss the implications of
hydrodynamics on the thermal Green's functions in $1+1$ dimensions.
In section 4, we discuss the details of how to isolate the gauge invariant fluctuation of the
dilaton and the graviton corresponding to the sound channel.
We then derive the dispersion relation of the quasi normal mode for the sound channel.
We also evaluate the two point functions of the stress tensor
components  from gravity and show that 
 the Lorentz structure  and the structure of its pole
agrees with that  predicted from general considerations of hydrodynamics in
$1+1$ dimensions.
In section 5, we show that the ratio of $\xi/s$ is $1/4\pi$ for both the D1-brane background and that
of the fundamental string.  This is done in two ways: we match the dissipative part of the 
quasi-normal mode for the sound channel in gravity 
to that expected from general hydrodynamic considerations and more directly 
by applying the Kubo formula for bulk viscosity in terms of stress tensor correlators.
Both methods yield the ratio of bulk viscosity to entropy density to be
$1/4\pi$.  We then show that this ratio continues to be $1/4\pi$ for the class
of field theories in $1+1$ dimensions dual to D1-branes at cones over Sasaki-Einstein manifolds.
 Appendix A.  shows that the constraints 
we impose on the  graviton and the dilaton perturbations  of the D1-brane background 
are consistent with their equations of motion. Appendix B.  contains proofs  of the  two identitites
which  are used in our derivation of the equation for the sound mode.

\section{Gauge/gravity duality for the D1-brane}

In this section, we briefly review the
statement of  gauge/gravity duality for the case of $N$ D1-branes.
This enables us to set the notations and conventions that we will use and also
to state the  bounds on temperatures for which 
the bulk viscosity evaluated holographically from the gravity 
background can be trusted.

In \cite{Itzhaki:1998dd}, it was argued that 
$SU(N)$ Yang-Mills with $16$ supercharges in $1+1$ dimensions 
at large $N$ is dual to the near horizon supergravity solution  of D1-branes.
The near horizon supergravity solution in the Einstein frame is given by
\begin{eqnarray}
\label{d1-soln}
ds_{10}^2 &=& H^{-\frac{3}{4}}(r) (
-dt^2 + dx_1^2) + H^{\frac{1}{4}}(r) ( dr^2 + r^2 d\Omega_7^2),
\\ \nonumber
e^{\phi(r)} &=&  H(r)^{\frac{1}{2}}, \\ \nonumber
* F_3^{RR} &=& {6} L^6 \omega_{S_7},
\end{eqnarray}
where 
\begin{equation}
\label{defH-L}
H(r) =  \left( \frac{L}{r}\right)^6, \qquad \hbox{and}\;
L ^6 = g_{YM}^2 2^6 \pi^3 N \alpha'^4,
\end{equation}
and $d\Omega_7^2$ refers to the metric on the unit 7-sphere and 
$\omega_{S_7}$ its volume form. Note that, we have dualised here the 
Ramond-Ramond 
charge of the D1-brane. 
The gravity description is valid in the energy domain
\begin{equation}
\label{domain-1}
g_{YM}N^{\frac{1}{6}} << U<< g_{YM} \sqrt{N}.
\end{equation}
Here $U$ sets the energy scale. Note that the Yang-Mills coupling in 
$1+1$ dimensions has the units of energy. 
For completeness, we mention that the background in \equ{d1-soln} is the solution of 
type IIB supergravity equations of motion 
in 10 dimensions obtained from the following 
action
\begin{equation}
\label{iibaction-1}
S_{IIB} = \frac{1}{16\pi G_{10}} \int d^{10} x 
\sqrt{-g} \left[ R(g) - \frac{1}{2} \partial_M\phi \partial^M\phi - \frac{1}{2 \cdot 3!}
e^{\phi} ( F_3^{(RR)})^2 \right].
\end{equation}
At both ends of the domain in \equ{domain-1}, the
curvatures of the  supergravity solution 
in \equ{d1-soln} grow and the solution breaks down. 
In the UV i.e. $U>>g_{YM} \sqrt{N}$, one can trust the perturbative 
description of Yang-Mills theory at any value of $N$.
However
in the domain 
\begin{equation}
\label{domain-2}
g_{YM} << U <<  g_{YM} N^{\frac{1}{6}},
\end{equation}
the dual description is given by the near horizon geometry of the 
fundamental string solution. 
This background is obtained by performing a S-duality transformation on 
the D-brane background in \equ{d1-soln}
\begin{eqnarray}
\label{f1-soln}
ds_{10}^2 &=& H^{-\frac{3}{4}}(r) (
-dt^2 + dx_1^2) + H^{\frac{1}{4}}(r) ( dr^2 + r^2 d\Omega_7^2),
\\ \nonumber
e^{\phi(r)} &=&  H(r)^{-\frac{1}{2}}, \\ \nonumber 
* F_3^{NS} &=& {6} L^6 \omega_{S_7}.
\end{eqnarray}
Note that the only changes in the background
compared to that of the D1-brane is 
$\phi\rightarrow -\phi$ and the Ramond-Ramond flux replaced by the 
Neveu-Schwarz flux on the 7-sphere. The above background is a solution to the equations of motion
from the following  action
\begin{equation}
\label{iibaction-2}
S_{IIB} = \frac{1}{16\pi G_{10}} \int d^{10} x 
\sqrt{-g} \left[ R(g) - \frac{1}{2} \partial_M\phi \partial^M\phi - \frac{1}{2\cdot 3!}
e^{-\phi} ( F_3^{(NS)})^2 \right].
\end{equation}
Unlike the case of the AdS/CFT duality, the supergravity solution in 
\equ{d1-soln} and \equ{f1-soln} is not asymptotically AdS${}_3$ but only 
conformal to AdS${}_3$ \cite{Boonstra:1998mp}. 
Finally, deep in the IR i.e. $U<< g_{YM}$, the valid description is given in terms of the 
conformal field theory on the orbifold $(R^{8})^N/S_N$ for any $N$. 
Thus both in the UV and in the IR, the $1+1$ dimensional super Yang-Mills
flows to a conformal field theory. It is only in the domain
$ g_{YM} << U<< g_{YM} \sqrt{N}$ and in the limit of large $N$, one has
a dual description in terms of a supergravity solution.

To study hydrodynamics of the D1-brane theory, we need to consider the 
theory at finite temperature. The dual description is given in terms of 
the near horizon geometry of the non-extremal D1-brane solution which 
is given by 
\begin{eqnarray}
\label{d1-solnne}
ds_{10}^2 &=& H^{-\frac{3}{4}}(r) (
-f(r) dt^2 + dx_1^2) + H^{\frac{1}{4}}(r) \left( \frac{dr^2}{f(r)} + r^2 d\Omega_7^2\right),
\\ \nonumber
e^{\phi(r)} &=&  H(r)^{\frac{1}{2}}, \\ \nonumber
F_7^{RR} &=& {6} L^6 \omega_{S_7},
\end{eqnarray}
where 
\begin{equation}
\label{defne}
f(r) = 1 - \left( \frac{r_0}{r}\right)^6.
\end{equation}
The temperature  and the entropy density of the D1-brane theory is related to the non-extremal 
parameter $r_0$ by
\begin{equation}
\label{tempne}
T = \frac{3 r_0^2}{2\pi L^3}, \qquad s = \frac{1}{ 4G_3}\left( \frac{r_0}{L}\right)^4 =
 \frac{2 \pi^4}{4! G_{10}} {r_0^4}{L^3}.
\end{equation}
We can convert the domain of the validity of the D1-brane solution to ranges in temperature by 
identifying
$U_0 = r_0/\alpha'$ with the  $U$ in \equ{domain-1}, using the defintion of $L$ in \equ{defH-L}
and the following relations.
\begin{equation}
 \label{holgraphic}
g_{YM}^2 = \frac{g_s }{2\pi\alpha'}, \qquad G_{10}  = 2^3 \pi^6 g_s^2 \alpha^{\prime 4}.
\end{equation}
We obtain 
\begin{equation}
\label{domain-11}
\sqrt{\lambda} N^{-\frac{2}{3}} << T << \sqrt{\lambda}.
\end{equation}
Here, we have defined the t'Hooft coupling $\lambda = g_{YM}^2 N$. 
It is now clear that for large $N$, this is a sufficiently large domain.
Now for $T<< \sqrt{\lambda} N^{-2/3}$, the holographic dual of the
Yang-Mills theory is given by
the non-extremal fundamental string solution.
\begin{eqnarray}
\label{f1-solnne}
ds_{10}^2 &=& H^{-\frac{3}{4}}(r) (
-f(r) dt^2 + dx_1^2) + H^{\frac{1}{4}}(r) \left( \frac{dr^2}{f(r)} + r^2 d\Omega_7^2\right),
\\ \nonumber
e^{\phi(r)} &=&  H(r)^{-\frac{1}{2}}, \\ \nonumber
F_7^{NS} &=& {6}{L^6} \omega_{S_7}.
\end{eqnarray}
Again writing the domain in \equ{domain-2} in terms of temperature and the 
't Hooft coupling, we obtain that the above 
solution can be trusted in the following temperature
range
\begin{equation}
\label{domain-22}
\sqrt{\lambda}N^{-1} << T<< \sqrt{\lambda} N^{-\frac{2}{3}}.
\end{equation}
To conclude, for very high temperatures $T>> \sqrt{\lambda}$ and for very low
temperatures $T<< \sqrt{\lambda}N^{-1}$, the Yang-Mills theory flows to a 
free conformal  field theory. 

In this paper we evaluate the bulk viscosity $\xi$ using the holographic description 
for temperatures in the regime $\sqrt{\lambda} N^{-1}<< T<< \sqrt{\lambda}$.
From the above discussion, we see that 
for $T>>\sqrt{\lambda}$ and $T<<\sqrt{\lambda}$,
the theory flows to a conformal field theory. Therefore,
we expect the bulk viscosity to vanish  in these domains. However, in the regimes \equ{domain-11} and 
\equ{domain-22}, we will see that we obtain a non-trivial value of the bulk viscosity.

\section{Hydrodynamics  and the sound mode in $1+1$ dimensions}

In this section, we will discuss generalities of relativistic hydrodynamics in 
$1+1$ dimensions. 
As mentioned in the introduction, 
hydrodynamics in $1+1$ dimensions is special due to the absence of
shear.
 We discuss here the constraints  of
conservation laws and hydrodynamics on the structure of the thermal Green's function of the
stress tensor in $1+1$ dimensions. 
We show that this implies that the only hydrodynamic mode   is longitudinal and we will determine its 
dispersion relation. 

\subsection{Lorentz structure of the correlators}

We show using translational invariance and conservation of the stress tensor that its
Green's function is entirely characterized by a single function in $1+1$ dimensions.
We define the retarded Green's function of the stress energy tensor to be
\bej{def2ptfn}
G_{\mu\nu,\alpha\beta} (x-y) = -i \theta(x^0 -y^0)
\langle [T_{\mu\nu} (x), T_{\alpha\beta}(y)] \rangle.
\eej
Making use of translation invariance of the state, we can define  the Fourier transform of the above correlator denoted as $G_{\mu\nu,\alpha\beta}(k)$.
It is symmetric by definition in indices ($\mu,\nu$) and ($\alpha,\beta$). Further more, we have the following symmetry due to CPT invariance.
\bej{cptsymm}
G_{\mu\nu,\alpha\beta}(k) = G_{\alpha\beta,\mu\nu}.
\eej
The conservation of the stress-energy tensor leads to the following Ward identity
\bej{conswiden}
k^\mu G_{\mu\nu,\alpha\beta }(k) =0.
\eej
This suggests a useful tensor which forms a basis to write down the correlator is
\bej{deften1}
P_{\mu\nu} = \eta_{\mu\nu} - \frac{k_\mu k_\nu}{k^2}.
\eej
Note that $k^\mu P_{\mu\nu} =0$.
If the states involved in the expectation value in \equ{def2ptfn} are
Lorentz invariant, a convenient decomposition is
to split the components into a part which contains the
trace $\eta^{\mu\nu}\eta ^{\a\b} G_{\mu\nu,\alpha\beta}$ and the traceless. This is given by
\bej{lorentzsplit1}
G_{\mu\nu,\alpha\beta}(k) = P_{\mu\nu}P_{\alpha\beta} G_{B}(k^2)
+ H_{\mu\nu, \alpha\beta} G_S(k^2), 
\eej
where
\bej{defhten}
H_{\mu\nu,\alpha\beta} = \frac{1}{2} ( P_{\mu\alpha}P_{\nu\beta} + P_{\mu\beta}P_{\nu\alpha}) - P_{\mu\nu}P_{\alpha\beta}.
\eej
Note that $\eta^{\mu\nu} H_{\mu\nu,\alpha\beta} =0$ and the
two tensors in \equ{lorentzsplit1} are orthogonal;
 $P_{\mu\nu}P_{\alpha\beta} H^{\mu\nu}_{\;\;, \alpha'\beta'} =0$.
At this stage, it seems neccessary  that one
needs  2 functions to characterize
the 2-point function of the stress energy tensor namely,  $G_B$ and $G_S$.
Now substituting explicitly the value of $k_{\mu} = (-\omega, q) $, we find that
\beaj{explictpten}
P_{tt} = \frac{q^2}{\omega^2 -q^2}, \qquad
P_{tx} = \frac{\omega q}{\omega^2 -q^2}, \qquad
P_{xx} = \frac{\omega^2}{\omega^2 - q^2}.
\eeaj
Using the above components of the tensor $P_{\mu\nu}$,
it is easy to see that all components of
$H_{\mu\nu,\alpha\beta}$ vanish.
Therefore, the two point function of the stress tensor in a $1+1$ dimensional theory
is entirely dependent on just one function $G_{B}(k^2)$.
Note that due to Lorentz invariance, $G_B$ is a
 function of the Lorentz invariant quantity namely, $k^2$.
When Lorentz invariance is broken in thermal field theory, one has rotational invariance only.
For this situation, it is convenient to use the spatial projection operator $P_{\mu\nu}^T$ which
is defined as
\bej{defpt}
P^T_{tt}= P^T_{ti} =P^T_{it} = 0, \qquad P_{ij} = \delta_{ij} - \frac{k_ik_j}{\vec k^2}.
\eej
But for $1+1$ dimensions, it is easy to see that
this tensor vanishes identically.  Therefore, $P_{\mu\nu}$ defined in
\equ{deften1} is purely longitudinal in this case.
The only other projection operator in $1+1$ dimensions
 which is symmetric and  constructed out the momenta  and $\eta_{\mu\nu}$ is the
\bej{project}
\tilde P_{\mu\nu} = \frac{k_\mu k_\nu}{k^2}.
\eej
But tensors constructed from the  above operator do not satisfy the Ward identity  \equ{conswiden}.
Thus when Lorentz invariance is broken, one only has  the following change. The arguments of the functions in $G_B$ and $G_S$
changes from the Lorentz invariant quantity $k^2$ to $(\omega, q)$. Thus the two point function can be
written as
\bej{lorentzsplit}
G_{\mu\nu,\alpha\beta}(\omega, q ) = P_{\mu\nu}P_{\alpha\beta} G_{B}(\omega, q).
\eej
Writing it explicitly, we obtain
\beaj{explicty2ptcomp}
G_{tt,tt} = \frac{q^4}{(\omega^2 - q^2)^2 } G_{B}(\omega, q) , \qquad
G_{tt, tx} = \frac{q^3\omega}{(\omega^2 - q^2)^2 }G_{B}(\omega, q) , \\ \nonumber
G_{tt, xx} = \frac{\omega^2 q^2}{(\omega^2 - q^2)^2 }G_{B}(\omega, q) , \qquad
G_{tx,tx} = \frac{\omega^2 q^2}{(\omega^2 - q^2)^2 }G_{B}(\omega, q) , \\ \nonumber
G_{tx,xx} = \frac{\omega^3 q}{(\omega^2 - q^2)^2 }G_{B}(\omega, q) , \qquad
G_{xx,xx} = \frac{\omega^4}{(\omega^2 - q^2)^2 }G_{B}(\omega, q).
\eeaj
Thus all components of the thermal Green's function of the stress tensor are determined by
a single function $G_B$.

\subsection{Poles in the correlators}

We show here that the function  $G_B$, which determines the
thermal Green's function  must exhibit a hydrodynamic singularity due to 
the propagation of sound. 
Using Lorentz invariance, the stress tensor of a fluid in  $1+1$ dimensions is given by
\cite{Landau}
\begin{equation}
 \label{fluidstress}
T^{\mu\nu} = (\epsilon +P) u^{\mu} u^{\nu} + P\eta^{\mu\nu} -\xi ( u^{\mu} u^\nu + \eta^{\mu\nu})\partial_\lambda u^\lambda,
\end{equation}
where $u^\mu$ is the 2-velocity with $u_\mu u^\mu =-1$
 and $\xi$ is the bulk viscosity.
To obtain the linearized hydrodynamic equations, 
consider small fluctuations from 
the rest frame of the fluid.
We then have the following
\begin{eqnarray}
 \label{restframe}
& & T^{00} = \epsilon + \delta T^{00},  \qquad T^{0x} = \delta T^{0x},  \qquad T^{xx} = P + \delta T^{xx}, 
 \\ \nonumber
& & u^0  = 1 , \qquad u^x = \delta u^x.
\end{eqnarray}
Note that  $u^0 =1$ up to the linear order due to the constraint $u^\mu u_\mu  =-1$.
From the form of the stress tensor in \equ{fluidstress}, we can obtain the spatial variation of the
velocity in terms of the stress tensor to linear order
\begin{equation}
 \label{variveloc}
\partial_x \delta u^x = \frac{\partial_x \delta T^{0x}}{\epsilon +P}.
\end{equation}
Substituting this for the velocity in $T^{xx}$, we obtain to the linear order
\begin{eqnarray}
 \label{txxlinear}
T^{xx} &=& P + \delta T^{xx}, \\ \nonumber
&=& P - \frac{\xi} {\epsilon +P}  \partial_x \delta T^{0x}.
\end{eqnarray}
The hydrodynamic equations are $\partial _\mu T^{\mu\nu} =0$. These reduce to
\begin{equation}
 \label{hydroeq}
\partial_0 T^{00} + \partial_x T^{x0} = 0, \quad \partial_0 T^{0x} +  \partial_x T^{xx} =0.
\end{equation}
Substituting the form of the linearized form of the stress tensor given in \equ{txxlinear}, we obtain
\begin{eqnarray}
 \label{hydroeq1}
\partial_0 \delta T^{00} + \partial_x \delta T^{0x} =0, \\ \nonumber
\partial_0 \delta T^{0x} + \frac{\partial P}{\partial \epsilon} \partial_x \delta T^{00}
- \frac{\xi}{\epsilon + P} \partial_x^2 \delta T^{0x}=0.
\end{eqnarray}
Here we have used the fact that the thermodynamic variable $P$ depends  only on the energy density  $\epsilon$ by
some equation of state.
Performing the Fourier transform of the above equations in position and time, we obtain the following
algebraic set of equations for $\delta T^{00}$ and $\delta T^{0x}$
\begin{eqnarray}
 \label{hydroeq2}
-\omega \delta T^{00} +q \delta T^{0x} =0, \\ \nonumber
-i\omega \delta T^{0x} + iq  v_s^2 \delta T^{00} + q^2 \frac{\xi}{\epsilon +P} \delta T^{0x} =0,
\end{eqnarray}
where we have defined the speed of sound $v_s$ as
\begin{equation}
 \label{soundspeed}
v_s^2 = \frac{\partial P}{\partial\epsilon}.
\end{equation}
Eliminating $\delta T^{00}$ using the first equation,  we obtain the following equation for
fluctuations in $\delta T^{0x}$.
\begin{equation}
 \label{heydoreq3}
\left(-i \omega ^2  + i q^2 v_s^2  + \frac{\xi}{\epsilon +P} \omega q ^2  \right) \delta T^{0x} =0.
\end{equation}
Therefore the fluctuation in $\delta T^{0x}$ obey the dispersion with $\omega$ given by
\begin{equation}
 \label{dispersionrel}
\omega^2 - v_s^2 q ^2 + i \frac{\xi}{\epsilon +P}\omega q^2  =0.
\end{equation}
Solving this to the leading order, we obtain the following dispersion relation 
for this longitudinal mode which we call the sound mode.
\begin{equation}
 \label{modes}
\omega = \pm v_sq  - i \frac{\xi}{2(\epsilon +P)} q^2.
\end{equation}
From the equations in \equ{txxlinear} and \equ{hydroeq2},  it can be seen that the remaining fluctuations also
obey the same dispersion relation.
This implies by the usual arguments  of linear response theory \cite{kadmartin}
 that  the two point function of the components of the stress tensor  has a pole at the above value of $\omega$.
Thus we find the function $G_B$ in the retarded correlation functions has a pole at \equ{modes}.

\section{The sound channel in gravity}
As discussed in the earlier section, the D1-brane theory admits a holographic gravity dual for 
the ranges in temperatures given in \equ{domain-11} and \equ{domain-22}. In this section, we 
first study the gravity solution of the D1-brane and isolate 
a diffeomorphism invariant perturbation which we identify as the longitudinal mode 
corresponding to the sound mode.  
From the analysis of its equation of motion
and by imposing quasi-normal mode boundary conditions, we derive its dispersion relation. We then evaluate the 
two point function of the 
components of the  stress tensor of the D1-brane theory holographically and confirm that it has the structure
predicted by the general properties of hydrodynamics in previous section.

To simplify our analysis, we first consistently truncate the 10 dimensional near horizon geometry of the 
D1-brane in \equ{d1-solnne} to 3 dimensions by dimensionally reducing on the 7-sphere using the 
following ansatz.
\begin{eqnarray}
 \label{s7truncate}
ds_{10}^2 &=& e^{-{14} B(r) } g_{\mu\nu}(x) dx^\mu dx^\nu  + e^{2B(r)} L^2 d\Omega_7^2, \\ \nonumber
&=& e^{-14 B(r)} \left( -c_T^2(r) dt^2 + c_X^2(r) dz^2 + c_R^2 dr^2 \right) + e^{2B(r) } L^2 dS_{X_7}^2.
\end{eqnarray}
Using this ansatz in the 10-dimensional supergravity equations of motion, one obtains 
a set of coupled differential equations for the fields $c_T(r), c_X(r), c_R(r), \phi(r)$ and 
$B(r)$.  It can be shown that on identifying 
\begin{equation}
 B(r) = - \frac{1}{24} \phi(r) 
\end{equation}
and keeping the Ramond-Ramond flux through the 7-sphere constant, 
one can obtain a consistent truncation of the 10-dimensional equations to effectively 3-dimensions
\cite{Mas:2007ng,Cvetic:2000dm}. 
The truncated set of equations of motion can be obtained from the following 
 Einstein-dilaton system in 3 dimensions with action
\be
\label{einstein-dil}
S=\f{1}{16\pi G_3}\int d^3x\sqrt{-g}\lf [R-\f{\b}{2}\pa _{\m}\p \pa ^{\m}\p-\cP(\p)\ri ].
\ee 
Here, $\b=\f{16}{9}$ and $G_3$ is a three dimensional Newton's constant. The dilaton is denoted by $\p$ and its potential is $\cP =-\f{24}{L^2}e^{\f{4}{3}\p}$. 
The coefficient in the dilaton potential is determined from the contributions due to the background flux through
the 7-sphere and from its curvature. 
The equations of motion are
\bea
R_{\m\n}&=&\f{\b}{2}\pa _{\m}\p\pa _{\n}\p +\cP(\p),\nn\\
\Box \p &=&\f{\cP ' (\p)}{\b}.
\eea
The D1-brane  in  10-dimension given in \equ{d1-solnne} reduces to 
\bea
\label{3dmetricsc}
ds^2&=&-c_T(r)^2dt^2+c_X(r)^2dz^2+c_R(r)^2dr^2,\\ \label{3ddilaton}
\p &=&-3\log\lf (\f{r}{L}\ri ),
\eea
with the components of the metric given by 
\be
\label{3dmetricco}
c_T^2=\lf (\f{r}{L}\ri )^8f,{\hspace{1.5 cm}}c_X^2=\lf (\f{r}{L}\ri )^8,{\hspace{1.5 cm}}c_R^2=\f{1}{f}\lf (\f{r}{L}\ri )^2,
\ee
with $f=1-\f{r_0^6}{r^6}$.
For future reference, we write down the 
the equations of motion explicitly in terms of these functions: 
\bea\label{En1}
\f{c_X''}{c_X}-\f{c_X'}{c_X}\f{c_R'}{c_R}+\f{\b}{4}{\p ^{\pr 2}}+\f{c_R^2\cP}{2}&=&0,\\\label{En2}
\f{c_T''}{c_T}-\f{c_T'}{c_T}\f{c_R'}{c_R}+\f{\b}{4}{\p ^{\pr 2}}+\f{c_R^2\cP}{2}&=&0,\\\label{En3}
c_R^2\cP+2\f{c_T'}{c_T}\f{c_X'}{c_X}-\f{\b}{2}\p ^{\pr 2}&=&0,\\\label{Dn1}
\p ''+\p '\ln '\lf (\f{c_Tc_X}{c_R}\ri )&=&\f{c_R^2\cP '}{\b}.
\eea

\subsection{Linearized equations of motion for the perturbations}

We consider very small wave like perturbations in the background of the above solution
 $g_{\m\n}\ra g_{\m\n}+\d g_{\m\n}$ and $\p \ra \p +\delta\phi$. 
Due to translational invariance along the D1-brane directions, we can assume that all 
perturbations can be expanded using  its Fourier mode. Focusing on one such mode, we have
\begin{equation}
 \delta g_{\mu\nu}( t, z, r) = e^{-i(\o t-qz)} h_{\mu\nu}(r), \quad
\delta\phi(t, z, r) = e^{-i(\o t-qz)} \varphi(r).
\end{equation}
We further  parametrize the metric perturbations as  
\begin{equation}
h_{tt}=c_T^2H_{tt}, \quad h_{tz}=c_X^2H_{tz}, \quad h_{zz}=c_X^2H_{zz}.
\end{equation}
Following \cite{Policastro:2002se,Policastro:2002tn,Kovtun:2005ev},
we  fix the gauge by choosing $\d g_{r\m}=0$.  
The equations of motion up to linear order in perturbations are
\beaj{dyneq1}
H_{tt}'' + \ln'\left(\frac{c_T^2 c_X}{c_R}\right) H_{tt}'
-\ln'(c_T) H_{zz}' - \frac{c_R^2}{c_T^2} Z_0 
- 2c_{R}^2 \frac{\partial {\cal P}}{\partial \phi} \varphi =0,
\\
\label{dyneq2}
H_{tz}'' + \ln'\left( \frac{c_X^3}{c_Tc_R} \right) H_{tz}' = 0,
\\
\label{dyneq3}
H_{zz}'' + \ln'\left( \frac{c_Tc_X^2}{c_R} \right) H_{zz}'
-
H_{tt}' \ln'(c_x) + \frac{c_R^2}{c_T^2} Z_0 
+ 2c_R^2 \frac{\partial{\cal P}}{\partial\phi} \varphi =0,
\\
\label{dyneq4}
\varphi'' + \ln'\frac{c_Tc_X}{c_R}\varphi' +
c_R^2 \left( \frac{\omega^2}{c_T^2} - \frac{q^2}{c_X^2} \right)\varphi +
\frac{1}{2}\phi'( H_{zz}' - H_{tt}')
-\frac{c_R^2}{\beta}\frac{\partial^2{\cal P} }{\partial\phi^2} \varphi =0,
\eeaj
where $A_t=q^2\f{c_T^2}{c_X^2}$ and $Z_0=A_tH_{tt}+2q\o H_{tz}+\o ^2 H_{zz}$. We also obtain following first order constraints from Einstein equations for $\d R_{r\m}$.
\beaj{constraint1}
H_{zz}' + \frac{q}{\omega} H_{tz}' - \frac{1}{\omega^2} A_t\ln'\frac{c_X}{c_T}
H_{tt} + \frac{1}{\omega^2} \ln'\frac{c_X}{c_T} Z_0
+ \beta\phi'\varphi =0,
\\
\label{constraint2}
H_{tt}' - \ln'\left( \frac{c_X}{c_T}\right)H_{tt} + \frac{\omega c_X^2}{q c_T^2} H_{tz}'
- \beta\phi'\varphi =0, \\
\label{constraint3}
\ln'(c_T) H_{zz}' - \ln'(c_X) H_{tt}'-\beta\phi'\varphi' +\frac{c_R^2}{c_T^2} Z_0
 + c_R^2 \frac{\partial{\cal P}}{\partial\phi}\varphi
=0.
\eeaj
In the appendix A, we show that the above 3 constaints can be consistently imposed on the 
4 dynamical equations of motion. 

\subsection{Diffeomorphism invariant sound mode}

Fixing the gauge $\delta g_{\mu r}=0$ does not exhaust all gauge degrees of freedom. 
One is left with the residual gauge freedom under the infinitesimal diffeomorphisms
$x^\mu \rightarrow x^\mu + \xi^\mu$ with  $\mu \in \{t, z, r\}$. 
Following the approach of \cite{Kovtun:2005ev},  
we will construct a quantity using the above perturbations which will be invariant under the diffeomorphism of the metric i.e. $\d g_{\m\n}\ra \d g_{\m\n}-\na _{\m}\xi _{\n}-\na _{\n}\xi _{\m}$.
The perturbations change under diffeomorphism as
\bea
\d g_{\m\n}&\ra &\d g_{\m\n}-\na _{\m}\xi _{\n}-\na _{\n}\xi _{\m},\nn\\
\d g_{tt}&\ra &\d g_{tt}-2\na _{t}\xi _{t}\nn
\ra \d g_{tt}+2i\o\xi _t+\f{(c_T^2)'}{c_R^2}\xi _r,\nn\\
\d g_{tz}&\ra &\d g_{tz}+i\o \xi _z+iq\xi _t,\nn\\
\d g_{zz}&\ra &\d g_{zz}-2iq \xi _z -\f{(c_X^2)'}{c_R^2}\xi _r.
\eea
Then, the combination $Z_0=A_tH_{tt}+2q\o H_{tz}+\o ^2H_{zz}$ changes as
\bea
Z_0&\ra& Z_0+\f{q^2}{c_X^2}(2i\o)\xi _t +\f{q^2}{c_X^2}\f{(c_T^2)'}{c_R^2}\xi _r+\f{2q\o}{c_X^2}(i\o \xi _z-iq\xi _t)-\f{\o ^2}{c_X^2}\lf (2iq\xi _z+\f{(c_X^2)'}{c_R^2}\xi _r\ri ),\nn\\
\d Z_0&=&\f{2\xi _rA_H\ln'(c_X)}{c_R^2}
\eea
where
\begin{equation}
\label{A-H}
A_H=A_t\f{\ln 'c_T}{\ln 'c_X}-\o ^2.
\end{equation}
The dilaton also changes under diffeomorphism as
\bea
\vp&\ra &\vp-\pa ^{\m}\p\xi _{\m}
\ra \vp -\f{\p '}{c_R^{2}}\xi _r.
\eea
We find the following combination gauge invariant.
\be\label{gauginv}
Z_P = Z_0 + A_\varphi\varphi {\hs{1.5 cm}}
{\tr{where}}\;\;\;A_{\varphi} = \frac{2 A_H \ln'(c_X) }{ \phi'}.
\ee
Note that unlike the case of higher dimensional
Dp-branes with  $p\geq 2$,  studied by \cite{Mas:2007ng}, there is only a single gauge invariant mode for
$p=1$.   Note that the $Z_0$ constructed in \cite{Mas:2007ng} for $p\geq 2$ cannot be trivially 
extended for this case as $H$ as defined in \cite{Mas:2007ng} does not exists for
$p=1$. Here, we note that the role of $H$ for $p=1$  is played by the dilaton fluctuation.
We will call the diffeomorphism invariant 
fluctuation $Z_p$ as the sound mode as it constitutes fluctuations longitudinal to the wave directions.

Next we outline the steps involved in obtaining the second order equation satisfied
by this mode.
 From the definition, we have
\beaj{doubderzp}
Z_P'' &=& A_t'' H_{tt} +2 A_t' H_{tt}' + A_{\varphi}''\varphi + 2A_{\varphi}'\varphi'
\\ \nonumber
& & + A_tH_{tt}'' + 2 q\omega H_{tz}'' + \omega^2 H_{zz}'' + A_\varphi\varphi''
\eeaj
where 
\begin{equation}
 A_t = q^2 \frac{c_T^2}{c_X^2}.
\end{equation}
Using  equations (\ref{dyneq1})-(\ref{constraint3}), we simplify above to
\bea
Z_P''&+&\ln '\lf (\f{c_Tc_X}{c_R}\ri )Z_P'-\f{c_R^2}{c_T^2}(A_t-\o ^2)Z_P\nn\\
&=&H_{tt}\lf [-4A_t\lf [\ln '\lf (\f{c_T}{c_X}\ri )\ri ]^2+A_t''+A_t'\ln '\lf (\f{c_Tc_X}{c_R}\ri )\ri ]\nn\\
&&+\vp\lf [4A_t\ln '\lf (\f{c_T}{c_X}\ri )\b\p ' +\ln '\lf (\f{c_Tc_X}{c_R}\ri )A_{\vp}'+A_{\vp}''+2c_R^2\cP '(A_t-\o ^2)+\f{c_R^2}{\b}\cP ''A_{\vp}\ri ]\nn\\
&&+2A_{\vp}'\vp'.
\eea
Since $A_t'=2A_t\ln '\lf (\f{c_T}{c_X}\ri )$,
\be
\label{id3-1}
A_t''=4A_t \lf [\ln '\lf (\f{c_T}{c_X}\ri )\ri ]^2+2A_t\ln ''\lf (\f{c_T}{c_X}\ri ).\ee
Evaluating the difference of the Einstein equations (\ref{En1})-(\ref{En2}), we obtain
\be
\label{id3}
\ln '\lf (\f{c_T}{c_X}\ri )\ln '\lf (\f{c_Tc_X}{c_R}\ri )+\ln ''\lf (\f{c_T}{c_X}\ri )=0.\ee
Using equation \equ{id3-1} and (\ref{id3}), we obtain
\bea
Z_P''&+&\ln '\lf (\f{c_Tc_X}{c_R}\ri )Z_P'-\f{c_R^2}{c_T^2}(A_t-\o ^2)Z_P\nn\\
&=&
\vp\lf [4A_t\ln '\lf (\f{c_T}{c_X}\ri )\b\p ' +\ln '\lf (\f{c_Tc_X}{c_R}\ri )A_{\vp}'+A_{\vp}''+2c_R^2\cP '(A_t-\o ^2)+\f{c_R^2}{\b}\cP ''A_{\vp}\ri ]\nn\\\label{LZP}
&&+
2A_{\vp}'\vp '.
\eea
Using Einstein equations and the 
dilaton equation of motion (\ref{En1})-(\ref{Dn1}), one can proove the following identities 
(see Appendix. \ref{2iden} ).
\bea
\label{ApApr}
\b\p '(A_t-\o ^2)+A_{\vp}'&=&-\lf [\f{\p ''}{\p '}+\ln '\lf (\f{c_X}{c_Tc_R}\ri )\ri ]A_{\vp},\\\label{Atphi}
4A_t\ln '\lf (\f{c_T}{c_X}\ri )\b\p ' +\ln '\lf (\f{c_Tc_X}{c_R}\ri )A_{\vp}'+A_{\vp}''&+&2c_R^2\cP' (A_t-\o ^2)+\f{c_R^2}{\b}\cP ''A_{\vp}\nn\\
&=&2A'_{\vp}\lf [\f{c_R'}{c_R}-\f{\p ''}{\p '}\ri ].
\eea
Using relation (\ref{ApApr}), one can write the relation (\ref{derzp}) as
\be\label{derzp2}
A_{\vp}\lf [\lf (\f{c_R'}{c_R}-\f{\p ''}{\p '}\ri )\vp +\vp '\ri ]=Z_P'+\ln '\lf (\f{c_X}{c_T}\ri )Z_P
\ee
Finally using relation (\ref{Atphi}), the equation (\ref{LZP}) can be written as
\be
Z_P''+\ln '\lf (\f{c_Tc_X}{c_R}\ri )Z_P'-\f{c_R^2}{c_T^2}(A_t-\o ^2)Z_P=2A_{\vp}'\lf [\vp '+\lf (\f{c_R'}{c_R}-\f{\p ''}{\p '}\ri )\vp\ri ].
\ee
Using the relation (\ref{derzp2}), we obtain
\be
\label{equationzp}
Z_P''+\lf [\ln '\lf (\f{c_Tc_X}{c_R}\ri )-2\f{A_{\vp}'}{A_{\vp}}\ri ]Z_P'-\lf [\f{c_R^2}{c_T^2}(A_t-\o ^2)+2\f{A_{\vp}'}{A_{\vp}}\ln '\lf (\f{c_X}{c_T}\ri )\ri ]Z_P=0.
\ee
This is the equation for  $Z_P$ from which we will obtain the dispersion relation 
for its quasi-normal mode. 
Here we make the following observation regarding the equation of $Z_P$.
Consider a minimally coupled scalar $\Psi$ in the background \equ{3dmetricsc}, 
its equation of motion is given by 
\begin{eqnarray}
 \frac{1}{\sqrt{-g}} \partial_\mu \left( \sqrt{-g} g^{\mu\nu} \partial_\nu \Psi \right) =0, \\ \nonumber
\psi'' + \ln'\left( \frac{c_Tc_X}{c_R}\right) \psi' - \frac{c_R^2}{c_T^2} (A_t -\omega^2) \psi =0,
\end{eqnarray}
where $g_{\mu\nu}$ is the background metric in \equ{3dmetricsc}.
The second line is obtained from the first by  focusing on the Fourier component 
$\Psi(t, z, r) = e^{-i(\omega t + qz)}$. Now comparing the above equation with the equation 
for $Z_P$ we see that the the equation for $Z_P$ \equ{equationzp}
is almost a equation for a minimally coupled
scalar except for the terms that are proportional to $A_{\varphi}'/A_{\varphi}$. 
Note that the dilaton dependence of the equation for $Z_P$ is entirely contained in 
these terms. 

Finally we emphasize that the analysis to obtain the equation for $Z_P$ in \equ{equationzp} depended only 
on the fact that the background is of the form \equ{3dmetricsc} with a radially dependent dilaton profile \equ{3ddilaton}
that satisfy the equations of motion \equ{En1}, \equ{En2}, \equ{En3} and \equ{Dn1}
obtained from the Lagrangian in \equ{einstein-dil}. At each step we have not assumed any form for the 
functions $c_{T}, c_{X}, c_R, \phi$,  the dilaton potential ${\cal P}$ and
any specific value for the constant $\beta$.
Therefore for any radially symmetric solution of the form \equ{3dmetricsc}
and a dilaton profile which solves the dilaton equation of motion, 
the equation  for $Z_P$  is  given by \equ{equationzp}. 

\subsection{The dispersion relation of the  sound mode}

For further analysis, we will use the explicit expressions of metric coefficients and dilaton.
We first  change the 
variable to $Y=\f{Z_P}{A_{\vp}}$ 
and the independent variable to $u=\f{r_0^2}{r^2}$, then the equation becomes
\be
\pa _{u}^2Y-\f{(2+u^3)}{u(1-u^3)}\pa _{u}Y-\lf [\f{q^2L^6}{4r_0^4}\f{(1-\l -u^3)}{(1-u^3)^2}+\f{18u^4(4\l -3)}{(1-u^3)(4-4\l -u^3)^2}\ri ]Y=0,
\ee
where $\lambda = \frac{\omega^2}{q^2}$. 
First we look for its solution near the horizon.
For $1-u\ll 1$, the equation becomes
\be
\pa_u ^2 Y-\f{1}{1-u}\pa _uY+\lf [\f{\a^2\o ^2 }{9(1-u)^2}+\f{6}{(3-4\l )(1-u)}\ri ]Y=0,
\ee
where $\a=\f{L^3}{2r_0^2}$.
To pick up the ingoing solution at the horizon, let us define $x= \ln(1-u)$. Then in terms of 
$x$ and for $1-u <<1$, the above equation reduces to the oscillator  equation
\begin{equation}
 \lf (\partial_x^2  + \frac{\a ^2\o ^2}{9}\ri ) Y =0.
\end{equation}
Therefore, the ingoing solution for $1-u<<1$ behaves as
  $e^{-\frac{i}{3} \a\o x} =  (1-u)^{-\frac{i}{3} \a\o }$. 
We next consider a solution of the type 
$$Y=(1-u^3)^{-\f{i}{3}\a\o}Z(u).$$ Then the
 equation for $Z(u)$ is given by
\bea
\pa _u^2 Z&-&\f{\pa _{u}Z}{u(1-u^3)}[2+u^3-2i\a\o u^3]\nn\\&&
+Z\lf [\a ^2\o ^2 \f{(1-u^4)}{(1-u^3)^2}-\f{\a ^2q ^2}{(1-u^3)}-\f{18u^4(4\l -3)}{(1-u^3)(4-4\l-u^3)^2}\ri ]=0.
\eea
In this equation, all terms are dimensionless. From the definition of temperature in \equ{tempne}, we see that 
$\alpha \sim 1/T$, therefore in the hydrodynamic limit
\begin{equation}
 \omega << T\;\;\;{\tr{and}}\;\;\; \quad q<<T,
\end{equation}
we ignore
 terms of order $q^2/T^2,\; \o ^2/T^2 ,\; \o q/T^2$ and higher, but keep terms of order $\o/T$, $q/T$ .
Performing this limit in the equation for $Z$, we obtain
\be
\pa _u ^2Z-\f{\{2+(1-2i\a\o )u^3\}}{u(1-u^3)}\pa _u Z
-\f{18u^4(4\l -3)}{(1-u^3)(4-4\l-u^3)^2}Z=0.
\ee
The well behaved solution of the above equation at $u=1$, the horizon,  is given by
\be
\label{finalsolnz}
Z=\f{6\l-2(1-\l)(3-4i\a\o )-u^3(3+2i\a\o )}
{12(3-2i\a\o )(4-4\l + u^3)}.
\ee
To obtain the quasi-normal modes of this solution, we need to impose the 
  Dirichlet condition $Z=0$ at the boundary at $u\rightarrow \infty$. 
This leads to the following  cubic equation for $\o$.
\be
-4i\a\o ^3+6 \o ^2+4i\a\o q^2-3q^2=0.
\ee
Solving this equation perturbatively by assuming  $\o \sim q$,
we obtain  the following dispersion relation for the sound mode.
\bej{dispersionrelg}
\o = \pm\frac{1}{\sqrt{2}} q - i \frac{\a }{6}q^2+...
\eej
with $$\alpha = \frac{L^3}{ 2 r_0^2} = \frac{3}{4\pi T}.$$

\subsection{Holographic evaluation of the stress tensor correlators}

In this section, we use the standard prescription of the gauge/gravity correspondence 
to evaluate the stress tensor correlations.  For this,
we  first need to expand the action in \equ{einstein-dil} along with the Gibbons-Hawking boundary term 
to second order in the fluctuation $H_{\mu\nu}$.  
The bulk action and the boundary term is given by 
\bea
\label{einstein-dil1}
S&=& S_{bulk} + S_{GH}, \\ \nonumber
S&=&\f{1}{16\pi G_3}\int d^3x\sqrt{-g}\lf [R-\f{\b}{2}\pa _{\m}\p \pa ^{\m}\p-\cP(\p)\ri ]
+ \frac{1}{8\pi G_3} \int d^2 x\sqrt{-h} K|_{r\rightarrow \infty}.
\eea
where
\[\b=\f{16}{9}\;\;\;\cP=-\f{24}{L^2}e^{4\p /3}.
\]
and $h$ is the boundary metric at a large but fixed value of $r$ and $K$ is the extrinsic 
curvature of the boundary \footnote{In general there are counter terms 
one has to add to regulate the action 
\cite{Cai:1999xg,Batrachenko:2004fd,Wiseman:2008qa,Kanitscheider:2008kd},
 these counter terms are crucial
to regulate the one point function of the stress tensor.
However for the two point functions which 
we will be interested in, they are not relevant.}.
Using the equations of motion \equ{dyneq1}-\equ{dyneq4} and the constraints 
\equ{constraint1}-\equ{constraint2} we can rewrite the 
expansion of the bulk action to second order in the fluctuations as a total 
derivative in the radial coordinate. This is given by
\bea
 S^{(2)}_{bulk}
&=&\int d^{3}x\lf [-\f{3c_X^3}{2c_Tc_R}H_{tz}H_{tz}'+\f{c_X^3}{c_Tc_R}\lf (\f{c_T'}{c_T}-2\f{c_X'}{c_X}\ri )H_{tz}^2-\f{\b}{2}
\f{c_Tc_X}{c_R}\vp \vp '
\ri .\nn\\&&
+\f{c_Tc_X}{4c_R}(H_{tt}+H_{zz})(H_{tt}+H_{zz})'+\f{c_Tc_X}{2c_R}\ln '(c_Tc_X)H_{tt}H_{zz}\nn\\&&
+\f{c_Tc_X}{4c_R}\lf \{H_{tt}H_{tt}'+H_{zz}H_{zz}'+(H_{tt}-H_{zz})\lf (\f{c_T'}{c_T}H_{tt}-\f{c_X'}{c_X}H_{zz}\ri )\ri \}\nn\\&&\lf .
+\f{\b}{4}\f{c_Tc_X}{c_R}(H_{tt}-H_{zz})\p '\vp
\ri ]'.\label{S26}
\eea
The second order perturbation in extrinsic curvature term is given by
\bea
\sqrt{-h}K&=&-\f{c_X^3}{8c_Tc_R}\lf [4\lf (\f{c_T'}{c_T}H_{tz}^2-3\f{c_X'}{c_X}H_{tz}^2-2H_{tz}'H_{tz}\ri )\ri .\nn\\
&&\lf .
+\f{c_T^2}{c_X^2}(H_{tt}+H_{zz})\lf (2(H_{tt}+H_{zz})'+\ln '(c_Tc_X)(H_{tt}+H_{zz})\ri )\ri ].
\eea
Since the bulk action at second order in perturbation is just a total derivative we just have 
to evaluate its contribution at the boundary.
Then the boundary action at second order in fluctuations  including the Gibbons Hawking term 
then reduces to 
\be
S^{(2)}=\f{1}{16\pi G_3}\int \f{d\o dq}{(2\pi )^2} \cL
\ee
where
\be
\cL =\f{c_X^3}{4\o q c_Tc_R}Z_P(\vk)H_{tz}' +\cL _{\rm{contact}},
\ee
and $\cL _{\rm{contact}}$ represents the part of the Lagrangian without any derivatives.
\bea
\cL _{\rm{contact}}&=&-\f{c_Tc_X}{4c_R}\lf [\ln '(c_Tc_X)H_{tt}H_{zz}+\f{c_X'}{c_X}H_{tt}^2+\f{c_T'}{c_T}H_{zz}^2-2\f{q}{\o }\ln '\lf (\f{c_X}{c_T}\ri )H_{tz}H_{tt}\ri ]\nn\\&&
+\f{c_X^3}{c_Tc_R}\f{c_X'}{c_X}H_{tz}^2-\f{c_Tc_X}{4c_R}\vp \lf [2\vp\lf \{\f{c_R^2\cP '}{\p '}-\b\ln '(c_Tc_X)\ri \}
-2\b \p '(H_{tt}-H_{zz})\ri .\nn\\&& \lf .
+\f{A_{\vp}}{\o ^2}\ln '\lf (\f{c_X}{c_T}\ri )H_{tt}+\f{2Z_0}{\p '}\lf \{\f{c_R^2}{c_T^2}-\f{\ln 'c_T}{\o ^2}\ln '\lf (\f{c_X}{c_T}\ri )\ri \}\ri ].
\eea
To obtain the above form of the action we have used the constraints (\ref{constraint1})-(\ref{constraint3})
to rewrite all derivative terms in $\cL$ in terms of the derivative $H_{tz}'$. 
We can further manipulate the boundary action and reduce it to the form
\be\label{action}
S^{(2)}=\f{1}{16\pi G_3}\int \f{d\o dq}{(2\pi )^2}\cA(\o ,q, r)Z_P'(r,\vk)Z_P(r,-\vk)+S_{CT}^{(2)},
\ee
where
\be
\cA(\o ,q, r )=-\f{\b}{2A_{\vp}^2}\f{c_Tc_X}{c_R}.
\ee
The contact term in equation (\ref{action}) is
\bea
S_{CT}^{(2)}&=&\f{1}{16\pi G_3}\int \f{d\o dq}{(2\pi )^2}\lf [\cL_{\rm{contact}}\ri ]
-\cA\lf [A_t'H_{tt} +\lf (2A_t+\f{A_{\vp}}{2\b\o ^2}\ri )\ln '\lf (\f{c_X}{c_T}\ri )H_{tt}
\ri .\nn\\&&\lf .
+\vp\lf \{A_{\vp}'+\b\p '(A_t-\o ^2)-A_{\vp}\lf (\ln '(c_R)-\f{\p ''}{\p '}\ri )\ri \}
\ri .\\&&\lf .
+Z_0\lf \{-\ln '\lf (\f{c_X}{c_T}\ri )+\f{A_{\vp}}{\b\p '}\f{c_R^2}{c_T^2}-\f{A_{\vp}}{\b\p '\o ^2}\ln '(c_T)\ln '\lf (\f{c_X}{c_T}\ri )\ri \}
\ri ]Z_P(r,-\vk). \nonumber 
\eea

Now that we have the boundary action, we can evaluate the stress-tensor correlators 
by using the standard rules of the AdS/CFT correspondence.
Note that in this case the geometry is not asymptotically AdS. However we expect the 
rules of the AdS/CFT correspondence to still be valid for this case since it can 
be related to the anti-de Sitter geometry up to a conformal factor 
\cite{Boonstra:1998mp,Kanitscheider:2008kd}. 
The boundary values of the fluctuation $H_{\mu\nu}$  couple to the stress tensor of the boundary 
theory  as in \cite{Policastro:2002tn}
\bea
S_{\rm{coupling}}
=\f{1}{2}\int d^4 x [H^0_{tt}T^{tt}+H^0_{zz}T^{zz}+2H^0_{tz}T^{tz}].
\eea
Here, the indices are raised, lowered and contracted using the flat
 metric $ds^2=-dt^2+dz^2$. The superscript $0$ indicates the fact that we are looking 
at the $r \rightarrow \infty$ or the  
 $u=\frac{r_0^2}{r^2}\ra 0$ limit of the corresponding bulk fields. 
Thus  these correspond to the boundary values of the fluctuations $H_{\mu\nu}$. Using the above coupling and 
the rules of the AdS/CFT correspondence summarized in the equation
\begin{equation}
 \langle \exp( i S_{\rm{coupling}})\rangle = \exp[ i S^{(2)}( H^0_{\mu\nu })],
\end{equation}
we can evaluate the various two point functions of the components of the stress tensor.
Consider  the retarded two point function  
$G_{tt, tt} = -i \langle [T_{tt}, T_{tt}] \rangle $, using the AdS/CFT prescription we obtain
\be
G_{tt,tt}=-4\f{\d S^{(2)}}{\d H^0_{tt}(\vk)\d H^0_{tt}(-\vk)}.
\ee
From the definition of $Z_P$, we note that its boundary value is related to the 
boundary values of the fluctuations by 
\be
Z^0 _P = q^2 H^0_{tt}+2\o q H^0_{tz}+\o ^2H^0 _{zz}+A^0_{\vp}\vp ^0.
\ee
From \equ{finalsolnz}, we see that the 
 expansion of  $Z_P$ near the boundary is given by 
\bea
Z_P&=& C(1+...)+Du^3(1+...),\nn\\
&=&Z^0_P\lf [1+... +\f{D}{C}\f{r_0^6}{r^6}(1+... )\ri ].
\eea
Here $C(\o ,q)$ and $D(\o ,q)$ are independent of $u$.
The ellipses denote higher powers in $u=\f{r_0^2}{r^2}$. 
Now substituting the above expansion for $Z_P$, in $S^{(2)}$ it is easy to see that relevant term
in the boundary action 
is the first term in \equ{action} which involves the derivative of $Z_P$. 
This results in the following expression for the two point function of the stress tensor.
\be
G_{tt,tt}=-4\f{\d S^{(2)}}{\d H^0_{tt}(\vk)\d H^0_{tt}(-\vk)}=-\f{1}{16\pi G_3}\f{q^4}{(\o ^2-q^2)^2}\f{6r_0^6}{L^7}\f{D}{C}.
\ee
Comparing the above holographic result with \equ{explicty2ptcomp} we see 
that the expected Lorentz factor for this correlator is reproduced. Furthermore we can read out
the holographic value of $G_B$ as 
\be
G_{B}(\o, q)=-\f{1}{16\pi G_3}\f{6r_0^6}{L^7}\f{D}{C}.
\ee
The poles in the Green's function are therefore same as the zeros of the factor $C$.
 The Dirichlet boundary condition for the mode $Z_P =0$ at the horizon is equivalent
to setting $C=0$ as noted for the case of backgrounds asymptotic
to AdS  which were studied in 
\cite{Kovtun:2005ev}.
Thus the poles in the two point function of the stress tensor which 
leads to the dispersion relation of the sound mode is  as given in equation (\ref{dispersionrelg}).
As a consistency check of our calculations, we have evaluated all the 
remaining two point functions of the components of the stress tensor.
In each case the expected Lorentz factor given in (\ref{explicty2ptcomp})
is reproduced with the same expression for 
 $G_B (\o , q)$.

\section{The ratio of bulk viscosity to entropy density}

In this section, we  use two methods to evaluate the 
ratio of bulk viscosity to entropy density $\xi/s$  of the D1-brane theory in the gravity 
regime which is valid for 
temperatures in the range $\sqrt{\lambda} N^{-\frac{2}{3}} << T << \sqrt{\lambda}$.
We use the dispersion relation of the sound mode to read out this ratio,  then as a cross check
we determine this ratio by evaluating the bulk viscosity directly using the Kubo's formula 
for the bulk viscosity in one
spatial dimensions. 
We then show that the ratio $\xi/s$ continues to be the same for the temperatures in the 
range $\sqrt{\lambda}N^{-1} << T<< \sqrt{\lambda} N^{-\frac{2}{3}}$ when the 
D1-brane background is replaced by the F1-string solution. 
In both these regimes, the ratio $$\frac{\xi}{s} = \frac{1}{4\pi}.$$
We also show that 
for gauge theories corresponding to D1-branes at
cones over  Sasaki-Einstein 7-manifolds the ratio continues to be $1/4\pi$. 

\subsection{$\xi/s$ using dispersion relation of the sound mode}

Using the dispersion relation of the sound mode in gravity
\equ{dispersionrelg}  and comparing it with the dispersion 
relation of the sound mode using general hydrodynamic consideration \equ{dispersionrel}, 
we deduce  the transport properties of the D1-brane matter at temperature $T$.
As we have seen in section 3, general hydrodynamics considerations give the following dispersion relation
for the sound mode
\begin{equation}
\label{genhydro}
 \omega = v_s - \frac{i}{2} \frac{1}{\epsilon + P} \xi q^2, \qquad v_s^2 = \frac{\partial P}{\partial
\epsilon}.
\end{equation}
Here, $v_s$ is the speed of sound in the medium  $P$, the pressure and $\epsilon$ the energy density.
Now comparing this with the dispersion relation obtained from gravity in \equ{dispersionrelg}, we can 
read out the speed of sound in the D1-brane matter as
\begin{equation}
\label{speedsound}
 v_s^2 =\frac{1}{2}.
\end{equation}
It is now easy to see that from the definition of the speed of sound $v_s = \frac{\partial P}{\partial \epsilon}$, we 
obtain the equation of state
$$\epsilon = 2P$$
for the D1-brane matter. 
Thus the mediums seem to behave as a conformal fluid in $2$ spatial dimensions just like that of the 
M2-brane.  The fact that the  thermodynamic properties of the D1-brane theory is similar to the 
M2-brane was noted earlier in \cite{Peet:1998wn} and \cite{Mateos:2007vn}. 
In \cite{Peet:1998wn}, it was noted that the entropy  density of the D1-brane behaves like that 
of the M2-brane \footnote{See below equation 2.17 in \cite{Peet:1998wn}}, while in 
\cite{Mateos:2007vn} the speed of sound in the D1-brane matter was evaluated  holographically 
just  using thermodynamics of the D1-brane and shown to be the same as given in \equ{speedsound}.
It will be interesting to understand this coincidence since the
 D1-brane gravity solution is certainly not of the M2-brane form. 

From comparing the dissipative part of the dispersion relation in \equ{dispersionrelg} and 
\equ{genhydro} we see that 
\be
\label{compareratio}
 \frac{\xi}{\epsilon +P} = \frac{\alpha}{3} = \frac{L^3}{ 6r_0^2} =\frac{1}{4\pi T}.
\ee
In the absence of chemical potentials, we have the following thermodynamic relation
\begin{equation}
\label{thermorel}
 \epsilon + P = Ts.
\end{equation}
Substituting the above equation in \equ{compareratio}, we find that 
\begin{equation}
\label{idealratio}
 \frac{\xi}{s} =  \frac{1}{4\pi}.
\end{equation}
Using \equ{tempne}
 and the definition of $L$  from \equ{defH-L}, the entropy density can be 
written in terms of the field theory variables and temperature 
as 
\begin{equation}
 s = \frac{2^4 \pi^{\frac{5}{2} } }{3^3} \frac{N^2 T^2}{\sqrt{\lambda}}.
\end{equation}
Then the bulk viscosity is given by
\begin{equation}
 \xi =  \frac{2^6 \pi^{\frac{7}{2} } }{3^3} \frac{N^2 T^2}{\sqrt{\lambda}}.
\end{equation}

Let us now compare the result in
\equ{idealratio}  with  the ratio of bulk viscosity to entropy density  for Dp-branes with $p\geq 2$.
Using the results of \cite{Mas:2007ng}   for Dp-branes with $p\geq 2$ we have
\be
\frac{\xi}{s}=\frac{\xi}{\eta}\frac{\eta}{s}=\frac{1}{4\pi}\frac{2(3-p)^2}{p(9-p)}.
\ee
Note that the above equation is valid $p\geq2$,  since the method used by \cite{Mas:2007ng} 
is valid only for $p\geq 2$. 
But using the result \equ{idealratio},
it is clear that the formula for $\xi/s$ obtained for $p\geq 2$ continues to hold for also $p=1$.

\subsection{ $\xi/s$ using the Kubo's formula}

As a cross check of our calculations, we use the Kubo's formula to obtain the ratio of 
bulk viscosity to entropy density for the D1-branes.  In $1+1$ dimensions, 
Kubo's formula for bulk viscosity is given by (see for instance in  \cite{kadmartin}) 
\bea
\label{kuboformula}
\xi&=&\lim_{\o \ra 0}\f{1}{\o}\int _{0}^{\infty}dt\int dz e^{i\o t}\lan [T_{zz}(x),T_{zz}(0)]\ran.\\
&=&\lim_{\o \ra 0}\f{i}{\o}G_{zz,zz}(\o ,q=0). \nonumber
\eea
As we have seen in section 4.4, 
the two point function of the stress tensor from gravity is given by
\be
G_{zz,zz}=\f{\o ^4}{(\o ^2-q^2)^2}G_B(\o ,q),{\hs{1.5 cm}}G_B=-\f{1}{16\pi G_3}\f{r_0^6}{L^7}\f{D}{C}.
\ee
Here, the 
coefficients C and D are related to the asymptotic expansion of the solution near the boundary at $u \rightarrow 0$
given by
\be
Z_P=C(1+...)+Du^3(1+...).
\ee
For the sound mode in \equ{finalsolnz}, we find that
\bea
C&=&\f{(-4i\a \o ^3+4i\a \o q^2+6\o ^2-3 q^2)}{9(2i\a \o -3)},\nn\\
D&=&\f{\o[9i\o q^2+8i\a ^2\o (\o ^2-q^2)^2-12 \a (2q^4-3q^2\o ^2+\o ^4)]}{54(3i+2\a \o)(q^2-\o ^2)}.
\eea
In limit $q\ra 0$, we obtain
\bea
C\ra \f{-2\o ^2}{9}{\hs{1.5 cm}}D\ra\f{-2i\a\o ^3}{27}{\hs{1.5 cm}}\f{D}{C}\ra\f{i\a\o}{3},
\eea
and hence
\be
G_{zz,zz}(\o ,0)=G_{B}(\o ,0)=-\f{1}{16\pi G_3}\f{6r_0^6}{L^7}\f{i\a \o}{3}.
\ee
Now using the  temperature and the entropy density of the D1-branes solution given in 
\equ{tempne}, we can write the two point function as 
\be
G_{zz,zz}(\o ,0)=-Ts\f{\a\o}{3}=-i\f{\o s}{4\pi}.
\ee
Substituting this result for the two point function into Kubo's formula in 
\equ{kuboformula}, we obtain
the ratio
\be
\frac{\xi}{s} =\frac{1}{4\pi}.
\ee
This agrees with that obtained from 
examining the dissipative part of the sound mode in \equ{idealratio} and therefore
is a consistency check on our calculations.

\subsection{Universality of the ratio $\xi/s$}

\vspace{.5cm}
\noindent
\emph{(i) $\xi/s$ for the fundamental string}
\vspace{.5cm}

We have seen that in the range of temperatures  $\sqrt{\lambda} N^{-\frac{2}{3}} << T << \sqrt{\lambda}$
for which the D1-brane solution can be trusted, the ratio $\xi/s$ is given by the ideal value
$1/4\pi$. We now show that the ratio continues to be 
$1/4\pi$ for the temperature range $\sqrt{\lambda}N^{-1} << T<< \sqrt{\lambda} N^{-\frac{2}{3}}$
for which the fundamental string solution can be trusted. 
From \equ{f1-solnne}, we see that the only difference between the F1-string solution 
and that of the D1-brane solution is that of the dilaton profile. In fact the F1-string solution
can be obtained from the D1-brane solution by replacing $\phi\rightarrow -\phi$. 
On compactifying to $3$ dimensions 
just as for the D1-brane case, one obtains a consistent truncation of the equations of motion. 
The equations of motion can be derived from the action in $3$ dimensions given in \equ{einstein-dil}
but with
\begin{equation}
 \beta = \frac{16}{9}, \quad \mbox{and}\;\; {\cal P} = -\frac{24}{L^2} e^{-\frac{4}{3 }\phi}.
\end{equation}
The F1-string solution in 10 dimension given in \equ{f1-solnne} reduces to 
\bea
\label{3dmetricscf}
ds^2&=&-c_T(r)^2dt^2+c_X(r)^2dz^2+c_R(r)^2dr^2,\\ \label{3ddilatonf}
\p &=&3\log\lf (\f{r}{L}\ri ).
\eea
with $c_T, c_X, c_R$ and $f$ given by \equ{3dmetricco} just as for the D1-brane case.
Note that the only change from the D1-brane case is that of the sign of the dilaton.

The analysis of the linearized perturbation and obtaining the equation for the sound mode
in section 4.1 and 4.2 just depended on the fact that the background is 
of the form in \equ{3dmetricscf} and radially symmetric. Thus the equation for the sound 
mode for the F1-string solution continues to be
\equ{equationzpf} which is given by
\be
\label{equationzpf}
Z_P''+\lf [\ln '\lf (\f{c_Tc_X}{c_R}\ri )-2\f{A_{\vp}'}{A_{\vp}}\ri ]Z_P'-\lf [\f{c_R^2}{c_T^2}(A_t-\o ^2)+2\f{A_{\vp}'}{A_{\vp}}\ln '\lf (\f{c_X}{c_T}\ri )\ri ]Z_P=0.
\ee
In the above equation, the only dependence on the dilaton profile is due to  the 
ratio $\frac{A_\varphi'}{A_{\varphi}}$. 
From the definition of $A_\varphi$ in \equ{gauginv}, we see that this ratio is given by
\begin{equation}
 \frac{A_{\varphi}'}{ A_\varphi} =
\frac{[A_H \ln'(c_X)]'}{ A_H \ln'(c_X)} - \frac{\phi''}{\phi'}.
\end{equation}
Since $A_H$ just depends on the components of the metric
\equ{A-H}, the only difference for the F1-string can 
arise from the ratio $\phi''/\phi'$.  But it is easy to see
from \equ{3dmetricscf},  that this ratio is same as that of the D1-brane case.
Therefore the ratio $\frac{A_\varphi'}{A_\varphi}$ for the F1-string remains identical to that 
of the D1-brane case. 
Thus, the equation of the sound mode $Z_P$ is the same for the F1-string. Now the rest of the analysis
to extract the dispersion relation for the sound mode proceeds identical to that of the D1-brane case.
This results in the ratio 
$$
\frac{\xi}{s} = \frac{1}{4\pi}
$$ for the F1-string background.
We have thus extended the evaluation of the bulk viscosity to the temperature range
$\sqrt{\lambda}N^{-1} << T<< \sqrt{\lambda}$ for the  $U(N)$  gauge theory with
16 supercharges in $1+1$ dimensions.

\vspace{.5cm}
\noindent
\emph{(ii) $\xi/s$ for D1-branes at  cones over Sasaki-Einstein 7-manifolds}
\vspace{.5cm}

Here we show that  the ratio of the bulk viscosity to entropy density for 
theories of D1-branes at a  cone over 
Sasaki-Einstein 7-manifolds is also given by $1/4\pi$ in the regimes given by \equ{regimes}.
The near horizon geometry of the corresponding supergravity solution is given by
\begin{eqnarray}
 \label{grav-se}
ds^2 &=& H^{-\frac{3}{4}}(r) \left( - f(r) dt^2 + dx_1^2\right) + H^{\frac{1}{4}} (r) \left( \frac{dr^2}{f(r)} +
r^2 dS^2_{X_7} \right), \\ \nonumber
e^{\phi(r)} &=& H(r)^{\frac{1}{2} }, \\ \nonumber
F_{7} &=& {6} L^{6} \omega_{X_7}.
\end{eqnarray}
where
\begin{equation}
 \label{defh-se}
H(r) = \left( \frac{L}{r}\right) ^{6}, \qquad f(r) = 1 - \left( \frac{r_0}{r}\right)^6.
\end{equation}
and $dS^2_{X_7}$ stands for the metric of the 7 dimensional Sasaki-Einstein manifold $X_7$, and
$\omega_{X_7}$ is its volume form.  Note that the only way this background differs from D1-brane at
flat space is by the replacement of the 7-sphere by the 7-dimensional Sasaki-Einstein manifold.
To be explicit here, we write down the metrics of  two 7-dimension Sasaki-Einstein manifolds known
as $Q^{1,1,1}$ and $M^{1,1,1}$ in the literature.
The metric of $Q^{1,1,1}$ is given by \cite{Page:1984ae}
\begin{eqnarray}
 \label{met-se1}
dS^2_{Q^{1,1,1}} =  \frac{1}{16}( d\psi - \sum_{i=1}^3 \cos\theta_i d\phi_i)^2 +
 \frac{1}{8} \sum_{i=1}^3( d\theta_i^2 + \sin^2\theta_i d\phi_i^2 ),
\end{eqnarray}
while the metric of $M^{1,1,1}$ is given by \cite{Page:1984ad}
\begin{eqnarray}
 \label{met-se2}
dS^2_{M^{1,1,1}} &=& \frac{1}{64} \left( d\tau + 3 \sin^2 \mu  \sigma_3 + 2 \cos\theta_2 d\phi_2 \right)^2  \\ \nonumber
& & + \frac{3}{4} \left( d\mu^2 + \frac{1}{4} \sin^2 \mu ( \sigma_1^2 + \sigma_2^2 +
\cos^2 \mu \sigma_3^2 ) \right) \\ \nonumber
& & +\frac{1}{8} ( d\theta_2^2 + \sin^2 \theta_2d\phi_2^2),
\end{eqnarray}
 where
\begin{equation}
 \label{defcoord-se}
\sigma_1 = d\theta_1, \qquad \sigma_2 = \sin\theta_1 d\phi_1, \qquad \sigma_3 = ( d\psi + \cos\theta_1 d\phi_1).
\end{equation}
Similar to the case of flat space we now dimensionally  reduce  to $3$ dimensions using the ansatz
\begin{eqnarray}
 \label{dim-red-se}
ds_{10}^2 &=& e^{-{14} B(r) } g_{\mu\nu}(x) dx^\mu dx^\nu  + e^{2B(r)} L^2 dS_{X_7}^2, \\ \nonumber
&=& e^{-14 B(r)} \left( -c_T^2(r) dt^2 + c_X^2(r) dx_1^2 + c_R^2 dr^2 \right) + e^{2B(r) } L^2 dS_{X_7}^2.
\end{eqnarray}
Again one can verify just as in the case of flat space the identification
\begin{equation}
 B(r) = - \frac{1}{12} \phi(r),
\end{equation}
provides a consistent reduction provided the 7-form flux through the Sasaki-Einstein manifold is held constant.
The equations of motion reduce to the Einstein-dilaton system in $3$ dimensions.
The coefficient of the
potential of the dilaton in this system is dependent on the
the Ricci-scalar of the Sasaki-Einstein manifold. For the metric $L^2 dS_{X_7}^2$ when $X_7$ is either
$Q^{1,1,1}$ or $M^{1,1,1}$
the Ricci-scalar is given by
\begin{equation}
 \label{ricc-scal-se}
R(X_7)  = \frac{42}{L^2}.
\end{equation}
This is same as that of the 7-sphere, which implies the coefficient of the dilaton potential will remain the
same in the effective action. 
Note that the Ricci-scalar for the general $Y^{(p,q)}$ Sasaki-Einstein 7-manifolds 
constructed in  
\cite{Gauntlett:2004hh} and recently studied in \cite{Martelli:2008rt} is also given by \equ{ricc-scal-se} \footnote{See below equation
2.6 of  \cite{Martelli:2008rt}.}. 
Taking all this into consideration, the effective action in
$3$-dimensions is given by
\begin{equation}
 \label{3d-action-se}
I= \frac{1}{ 16\pi \tilde G_3} \int d^3 x \sqrt{-g} \left(
R(g) - \frac{8}{9} \partial_\mu \phi \partial^\mu\phi + \frac{24}{L^2} e^{\frac{4}{3}\phi } \right),
\end{equation}
where
\begin{equation}
 \frac{1}{\tilde G_3} = \frac{L^7 {\rm Vol} ( X_7)}{G_{10}}.
\end{equation}
We see apart from the definition of the 3 dimensional Newton's constant,  the effective action remains
the same as that of D1-branes in flat space.
The background values of $c_T(r), c_X(r), c_R(r)$ and $\phi(r)$ is also same as that of D1-branes in flat space.
Therefore the quasi-normal mode analysis remains identical and the ratio of the bulk viscosity to the entropy
density remains the same. 
The analysis for the case of the F1-strings at cones over
Sasaki-Einstein 7-manifolds is also identical to that of F1-strings in flat space and yields the same ratio
of bulk viscosity by entropy density.

The gauge theory of D1-branes at cones over Sasaki-Einstein 7-manifolds is different from that
of the $SU(N)$ theory with 16 supercharges for the case of D1-branes in flat space.
The number of supersymmetries and the matter content of the theory is different. 
Inspite of this for the temperature range \equ{regimes}, the theories of 
D1-branes at cones over Sasaki-Einstein 7-manifolds  have the 
same value of
$\xi/s =1/4\pi$ as that of D1-branes in flat space.

\section{Conclusions}

We have shown that the ratio of bulk viscosity to the entropy density for the 
 $SU(N)$ gauge theory  with 16 supercharges in $1+1$ 
dimensions on the D1-branes in the temperature range 
$\sqrt{\lambda}N^{-1} << T<< \sqrt{\lambda}$ is given by $1/4\pi$. 
For temperatures outside this range, the D1-brane gauge theory flows to a conformal field theory.
We therefore expect the bulk viscosity to entropy density of this theory to vanish for
$T<< \sqrt{\lambda} N^{-1}$ and $T>>\sqrt{\lambda}$.
 A technical result of our analysis is the equation for  the sound mode given in \equ{equationzp} for 
any Einstein-dilaton system in 3 dimensions given by the 
Lagrangian \equ{einstein-dil}  which admits a radially symmetric
solution of the  form \equ{3dmetricsc} and a dilaton profile determined by the 
dilaton equation of motion. 

We have also seen that for the theory of D1-branes at cones over Sasaki-Einstein 7-manifolds
the near horizon geometry dimensionally reduced to 3-dimensions is identical to that of 
D1-branes in flat space. This implies that the ratio of bulk viscosity to entropy density is given by 
$1/4\pi$. At this point, it is perhaps worthwhile to investigate other $1+1$ dimensional systems 
which admit gravity duals with 
different near horizon geometry.
 There are two possible interesting geometries one could explore:
One can turn on a R-charge along the Cartan directions of the $SO(8)$ R-symmetry of the D1-brane
theory. This results in a R-charged black hole in the Einstein-dilaton-Maxwell system in 
3-dimensions. 
The near horizon geometry of such black holes in 3-dimensions is different and it will be interesting
to  evaluate
the ratio of bulk viscosity to entropy density in this geometry. 
Another interesting geometry in 3-dimensions is that of the D1-D5 system. 
This geometry is conformal, however one could turn on a relevant operator in  the 
orbifold theory of the D1-D5 system
which will render it non-conformal.  The holographic dual 
 geometry to such  a system will necessarily have 
a non-trivial dilaton profile and therefore a non-trivial bulk viscosity.
This system will fall into the general Einstein-dilaton system studied in this paper, 
with an specific dilaton potential. We expect the dual geometry to be  radially symmetric 
since one can possibly choose the relevant deformation to be translationally invariant 
along the brane directions.
From the analysis in this paper we can then conclude that the equation for the sound mode is 
given by \equ{equationzp}.
It is tempting to speculate that the ratio of bulk viscosity to entropy density for this system 
will also by $1/4\pi$.  It will be interesting to construct this dual geometry for the
deformed D1-D5 system explicitly and verify this conjecture.

One of our motivations to study holographic duals to hydrodynamics in $1+1$ dimensions is the 
possibility of taking the non-relativisitc limit to obtain the Burger's equation
\footnote{Burger's equation is the non-relativistic Navier-Stokes equation
in one spatial dimension given by  
$\partial_t v + v \partial_x v = \xi \partial_x^2 v$, where $v$ is the velocity.}
 and study turbulence in 
one spatial dimension. However 
the scaling put forward in \cite{Fouxon:2008tb,Bhattacharyya:2008kq} to obtain the non-relativisitc limit
results in velocity of sound being infinite. 
This leads to  hydrodynamics with divergenceless velocity flow,
$$\partial_i v^i =0.$$
 In one spatial dimension,  this implies a trivial solution for the velocity field. 
The velocity field is constant in both space and time. 
It will be interesting to explore the possibility of  taking other  non-relativistic limits which can  
lead to the Burger's equation. 

\vspace{.5cm}
\noindent
{\bf{Note added:}}
After completion of this work, we noticed the preprint
 \cite{Kanitscheider:2009as} which has some overlap with this paper.

\acknowledgments
We would like to especially  thank Shiraz Minwalla  
for useful suggestions and insights at various stages in this project. 
We also thank  Rajesh Gopakumar and Nemani Suryanarayana for discussions. 
J.R.D thanks the organizers of the Monsoon Workshop on String Theory (2008) and ICTS, TIFR
for the stimulating environment which  resulted in  this project. 
M.M thanks the Centre for High Energy Physics, IISc. for hospitality during which part of the work was done.
Finallly we would like to thank the people of India for supporting research in fundamental physics.

\appendix
\section{Consistency of the constraint equations}

Note that the equations \equ{constraint1}-\equ{constraint3} form 3 constraints for the  4 dynamical equations
\equ{dyneq1}-\equ{dyneq4}.
These constraints must therefore be consistent with the dynamical equations. That is,
on evolving the constraints by the dynamical equations, one should not generate new constraints.
In this appendix, we show that by differentiating the 3 constraints in \equ{constraint1} to 
\equ{constraint3}  and reducing them again
to first order equations using the dynamical equations in \equ{dyneq1} to \equ{dyneq4}, one does not generate any 
new constraints. 
This allows us to conclude that the constraints are consistent with the dynamical equations and 
they reduce the number of dynamical variables to just one, which we have identified as the 
sound mode $Z_P$.  

We start  with writing  down 2 equations which we will use in our analysis. We define
\be
Z_0=A_tH_{tt}+2q\o H_{tz}+\o ^2H_{zz},
\ee
where $A_t=q^2\f{c_T^2}{c_X^2}$.
Its first derivative is
\be
Z_0'=A_t'H_{tt}+A_tH_{tt}'+2\o qH_{tz}'+\o ^2H_{zz}'.
\ee
Using the constraint equations, one can remove $H_{tt}'$ and $H_{zz}'$ in favour of $H_{tz}'$. Then the expression reduces to
\bej{derz0}
Z_0'=\lf \{A_t'+A_t\ln '\lf (\f{c_X}{c_T}\ri )\ri \}H_{tt}-\o ^2\ln '\lf (\f{c_X}{c_T}\ri )H_{zz}-2\o q\ln '\lf (\f{c_X}{c_T}\ri )H_{tz}+(A_t-\o ^2)\b\p '\vp .
\eej
Next we write down an equation obtained by differentiating $Z_P$.
\bea
Z_{P}' &=& Z_0' + A_\varphi\varphi'+A_{\vp}'\vp ,\nn \\ \label{derzp}
 &=& -\ln'\left(\frac{c_X}{c_T} \right) Z_P + \left\{ \ln'\lf (\frac{c_X}{c_T}\ri )A_{\varphi}
+( A_t -\omega^2)\beta\phi'+A_{\vp}'\right\} \varphi + A_{\varphi}\varphi'.
\eea
Here we have used the equation \equ{derz0}.
Equation (\ref{derzp}) can also be obtained by adding $A_t$ times (\ref{constraint1}) and $\o ^2$ times (\ref{constraint2}).

To check the consistency of the constraint equations, we first differentiate them, then substitute the dynamical equations and reduce the equation to an equation 
with at the most first derivatives. We then show that we do not obtain any new constraints.
We start with the equation \equ{constraint2}.
We get
\beaj{neweq1}
\nonumber
 & &-\ln'\left(\frac{c_Tc_X^2}{c_R} \right) H_{tt}'
+ H_{zz}' \ln'(c_T) + \frac{\omega c_X^2}{qc_{T^2}}\ln'\left( \frac{c_R}{c_Xc_T} \right)H_{tz}'
- \ln''\lf ( \frac{c_X}{c_T}\ri ) H_{tt} + \frac{c_R^2}{c_T^2} Z_P
\\
& &+ \left( 2 c_R^2 \frac{\partial{\cal P}}{\partial \phi} - \frac{c_R^2}{c_T^2} A_{\varphi} -\beta\phi'' \right)\varphi - \beta\phi'\varphi' =0.
\eeaj
When we use (\ref{id3}) on the combination of equations (\ref{neweq1})-(\ref{constraint3})+$\ln '\lf (\f{c_Tc_X}{c_R}\ri )$ times (\ref{constraint2}), we get
\be\label{r1}
\O \vp=0,
\ee
where
\be
\O= c_R^2\cP '-\ln '\lf (\f{c_Tc_X}{c_R}\ri )\b\p '-\b\p ''.
\ee
Now \equ{r1} is consistent for any fluctuation $\varphi$ because $\Omega =0$, 
 since the  back ground statisfies the  dilaton equation of motion (\ref{Dn1}).
Thus the equation (\ref{neweq1}) is consistent with equations of motion.
\par
Next we differentiate \equ{constraint1} and use the dynamical equations to remove
the second derivatives. Doing this, we get the equation
\beaj{neweq2}
\nonumber
& & - \ln'\left(\frac{c_Tc_X^2}{c_R} \right )H_{zz}' -\frac{q}{\omega}
\ln'\left(
\frac{c_X^3}{c_Tc_R} \right) H_{tz}'
+ \left\{ \ln'(c_X) - \frac{A_t}{\omega^2} \ln'\left(\frac{c_X}{c_T} \right)
\right \}H_{tt}'\nn\\&&
-\frac{1}{\omega^2} \left( A_t \ln'\frac{c_X}{c_T} \right)' H_{tt}
+\left\{ \frac{1}{\omega^2} \ln''\lf (\frac{c_X}{c_T}\ri )
 - \frac{c_R^2}{c_T^2} \right\} Z_P
+ \frac{1}{\omega^2} \ln'\lf ( \frac{c_X}{c_T}\ri ) Z_P'
\nn\\
& &
+  \left[
-\frac{1}{\omega^2}  \left\{ A_\varphi \ln'\lf (\frac{c_X}{c_T}\ri  )\right\}^{\prime}
+\beta\phi'' +
\frac{c_R^2}{c_T^2} A_{\varphi} - 2c_R^2\frac{\partial{\cal P}}{\partial \phi}
\right] \varphi\nn\\&&
+ \left\{ - \frac{1}{\omega^2} A_{\varphi} \ln \lf(\frac{c_X}{c_T}\ri ) + \beta\phi' \right\}
\varphi' =0.
\eeaj
Note that using equation (\ref{id3}), one can show that
\be\label{id5}
\ln ''\lf (\f{c_X}{c_T}\ri )-\lf [\ln '\lf (\f{c_X}{c_T}\ri )\ri ]^2=\ln '\lf (\f{c_T}{c_X}\ri )\ln '\lf (\f{c_X^2}{c_R}\ri ).
\ee
Using (\ref{id5}) on the combination of equations (\ref{neweq2}) -(\ref{constraint3})
  +$\f{A_t}{\o ^2}\ln '\lf (\f{c_X}{c_T}\ri )$ times
(\ref{constraint2}) +$\ln '\lf (\f{c_X^2}{c_R}\ri )$ times (\ref{constraint1}), one obtains
\be
-\O\vp=0.
\ee
Hence, the equation (\ref{neweq2}) can be written as a linear combination of earlier constraint equations.


The derivative of third constraint equation (\ref{constraint3}) gives us
\bea
& &H_{zz}'\lf [\ln ''(c_T)-\ln'c_T \ln ' \lf (\f{c_Tc_X^2}{c_R}\ri )-\ln 'c_X\ln 'c_T +\f{\b}{2}{\p '} ^2\ri ]\nn\\
& & +H_{tt}'\lf [-\ln ''(c_X)+\ln ' c_X\ln ' \lf (\f{c_T^2c_X}{c_R}\ri )+\ln 'c_X\ln 'c_T -\f{\b}{2}{\p ' }^2\ri ]\nn\\
& &+\vp '\lf [\b\p '\ln'\lf (\f{c_Tc_X}{c_R}\ri )-\b\p ''-\f{c_R^2}{c_T^2}A_{\vp}
+c_R^2\cP ' \ri ]\nn\\\label{dc3}
&  &+c_R^2\vp\lf [-2\cP '\ln '\lf (\f{c_Tc_X}{c_R}\ri )+\b \p '\lf (\f{\o ^2}{c_T^2}-\f{q^2}{c_X^2}\ri )+\f{A_{\vp}}{c_T^2}\ln '\lf (\f{c_T^3c_X}{c_R^2}\ri )-\f{A_{\vp}'}{c_T^2}\ri ]\nn\\& &
+\f{c_R^2}{c_T^2}\lf \{Z_P'
-\ln '\lf(\f{c_T^3c_X}{c_R^2}\ri ) Z_P\ri \} =0.
\eea
We use equation $\O =0$ to remove the $\b\phi ''\vp '$ term and equation (\ref{derzp}) to remove $Z_P'$. Next we add $2\ln '\lf (\f{c_Tc_X}{c_R}\ri )$ times (\ref{constraint3}) to it. We note that equation (\ref{id3}) can also be written as
\be\label{r2a}
\ln '' c_T+\ln 'c_T\ln '\lf (\f{c_Tc_X}{c_R}\ri )=\ln ''c_X +\ln 'c_X\ln '\lf (\f{c_Tc_X}{c_R}\ri ).
\ee
Using this equation with the above manipulations, we obtain
\be\label{r2}
\lf [\ln '' c_T+\ln 'c_T\ln '\lf (\f{c_Tc_X}{c_R}\ri ) -2\ln 'c_T\ln 'c_X+\f{\b}{2}{\p '}^2 \ri ](H_{zz}-H_{tt})'=0.
\ee
The term in the square bracket is a combination of Einstein equations of motion, namely equation (\ref{En2})-$\f{1}{2}$(\ref{En3}), hence it vanishes. This implies that \equ{dc3} can be written as a linear
combination of the three basic constraints upto equations of motion.

\section{Identities from background field equations}\label{2iden}

\vspace{.5cm}
\emph{ (i) Derivation of equation \equ{ApApr} }
\vspace{.5cm}

First we will derive equation (\ref{ApApr}) from the background equations of motion.
The ($\o ^2$) and ($q^2$) part of the equation should independently vanish.
Since
\be
A_{\vp}'=-\f{\p ''}{\p '}A_{\vp}+\f{2}{\p '}\lf [A_t\lf \{\ln ''c_T+2\ln 'c_T\ln '\lf (\f{c_T}{c_X}\ri )\ri \}-\o ^2\ln ''c_X\ri ],
\ee
  we need to show
\bea\label{Apr1}
\f{2}{\p '}\lf [\f{\b\p '^2}{2}+2\ln 'c_T\ln '\lf (\f{c_T}{c_X}\ri )+\ln ''c_T+\ln 'c_T\ln '\lf (\f{c_X}{c_Tc_R}\ri )\ri ]=0\\\label{Apr2}{\tr{and}}{\hs{2 cm}}
\f{2}{\p '}\lf [\f{\b\p '^2}{2}+\ln ''c_X+\ln 'c_X\ln '\lf (\f{c_X}{c_Tc_R}\ri )\ri ]=0.
\eea
Using equations (\ref{En2}) and (\ref{En1}), we obtain
\bea
\label{lnct2}
\ln ''c_T&=&\ln 'c_T\ln '\lf (\f{c_R}{c_T}\ri )-\f{\b}{4}\p ^{\pr 2}-\f{c_R^2}{2}\cP ,\\
\label{lncx2}
\ln ''c_X&=&\ln 'c_X\ln '\lf (\f{c_R}{c_X}\ri )-\f{\b}{4}\p ^{\pr 2}-\f{c_R^2}{2}\cP .
\eea
Replacing $\ln ''c_T(\ln ''c_X)$ in equation [\ref{Apr1}]([\ref{Apr2}]), we find that the left hand side vanishes due to equation (\ref{En3}).

\vspace{.5cm}
\noindent
\emph{(ii)  Derivation of \equ{Atphi} }
\vspace{.5cm}

Next, we will derive relation (\ref{Atphi}).
The ($\o ^2$) and ($q^2$) part of the equation should be independently satisfied.
They are
\bea\label{Atphi1}
G'+2c_R^2\cP '+2\f{c_R^2\cP''}{\b\p '}\ln 'c_X+G\lf [\f{2\p ''}{\p '}+\ln '\lf (\f{c_Tc_X}{c_R^3}\ri )\ri ]=0,\\
\label{Atphi2}
4\b\p '\ln '\lf (\f{c_T}{c_X}\ri )+K'+2c_R^2\cP '+\f{2c_R^2\cP ''}{\b\p '}\ln 'c_T+K\ln '\lf (\f{c_T^3}{c_Xc_R^3}\ri )+\f{2\p ''}{\p '}K=0.
\eea
where
\bea
G&=&\f{2}{\p '}\lf [\ln '' c_X-\f{\p ''}{\p '}\ln 'c_X\ri ],\\
K&=&\f{2}{\p '}\lf [\ln '' c_T+2\ln 'c_T\ln '\lf (\f{c_T}{c_X}\ri )-\f{\p ''}{\p '}\ln 'c_T\ri ].
\eea
To show (\ref{Atphi1}), we first evaluate $G'$.
\be
G'=-\f{2\p ''}{\p '}G+\f{2}{\p '}\lf [\ln ^{(3)}c_X-\f{\p ^{(3)}}{\p '}\ln 'c_X\ri ].
\ee
Using equations (\ref{En1}) and (\ref{Dn1}), we can write
\bea
\ln ^{(3)}c_X-\f{\p ^{(3)}}{\p '}\ln 'c_X
&=&\ln ''c_X\ln '\lf (\f{c_R}{c_X}\ri )
+\ln 'c_X\ln ''c_T
+\f{\p ''}{\p '}\ln 'c_X\ln '\lf (\f{c_Tc_X}{c_R}\ri )
\nn\\&&
-\f{2c_R^2\cP '}{\b\p '}\f{c_X'}{c_X}\f{c_R'}{c_R}
-\f{c_R^2\cP ''}{\b}\ln 'c_X-c_R^2\cP\ln 'c_R\nn\\&&
+\f{\p '}{2}\lf \{-2c_R^2\cP '+
\b\p '\ln '\lf (\f{c_Tc_X}{c_R}\ri )\ri \}.
\eea
Then the left hand side of equation (\ref{Atphi1}) becomes
\bea
LHS&=&G\ln '\lf (\f{c_Tc_X}{c_R^3}\ri )+\f{2}{\p '}\lf [\f{\b\p ^{\pr 2}}{2}\ln '\lf (\f{c_Tc_X}{c_R}\ri )+\ln ''c_X\ln '\lf (\f{c_R}{c_X}\ri )+\ln 'c_X\ln ''c_T
\ri .\nn\\&&\lf .
+\f{\p ''}{\p '}\ln 'c_X\ln '\lf (\f{c_Tc_X}{c_R}\ri )-\f{2c_R^2\cP '}{\b\p '}\f{c_R'}{c_R}\f{c_X'}{c_X}-c_R^2\cP\ln 'c_R
\ri ]\nn\\
&=&\f{2}{\p '}\lf [\f{\b\p ^{\pr 2}}{2}\ln '\lf (\f{c_Tc_X}{c_R}\ri )+\ln ''c_X\ln '\lf (\f{c_T}{c_R^2}\ri )+\ln 'c_X\ln ''c_T
\ri .\nn\\&&\lf .
+\f{2\p ''}{\p '}\ln 'c_X\ln 'c_R-\f{2c_R^2\cP '}{\b\p '}\f{c_R'}{c_R}\f{c_X'}{c_X}-c_R^2\cP\ln 'c_R\ri ].
\eea
Using equations (\ref{En3}) and (\ref{Dn1}), one can show that
\bea\label{ctcxpr}
-(\ln ''c_T\ln 'c_X+\ln 'c_T\ln ''c_X)&=&\f{1}{2}\lf [c_R^2\cP -\f{\b}{2}\p ^{\pr 2}\ri ]'\nn\\
&=&c_R^2P\ln 'c_R+\f{\b \p ^{\pr 2}}{2}\ln '\lf (\f{c_Tc_X}{c_R}\ri ).
\eea
Then we obtain
\bea
LHS&=&\f{2\ln 'c_R}{\p '}\lf [-2\ln ''c_X
+\f{2\p ''}{\p '}\ln 'c_X-\f{2c_R^2\cP '}{\b\p '}\f{c_X'}{c_X}-2c_R^2\cP \ri ].
\eea
Using equation (\ref{lncx2}) to replace $\ln ''c_X$ and equation (\ref{Dn1}) to replace $\p ''$, the $LHS$ then vanishes due to equation (\ref{En3}).

\par
To show (\ref{Atphi2}), we first evaluate
\bea
K'&=&-2\f{\p ''}{\p '}K
+\f{2}{\p '}\lf [\ln ''c_T\lf \{\ln '\lf (\f{c_R}{c_T^2}\ri ) +2\ln '\lf (\f{c_T}{c_X}\ri )
\ri \}+\f{c_T'}{c_T}\f{c_R'}{c_R}+\f{\b\p ^{\pr 2}}{2}\ln '\lf (\f{c_Tc_X}{c_R}\ri )
\ri .\nn\\&&
-c_R^2\cP '\p '-c_R^2\cP \ln ' c_R+2\ln 'c_T\ln ''\lf (\f{c_T}{c_X}\ri )+\f{2\p ''}{\p '}\ln 'c_T\ln '\lf (\f{c_T}{c_X}\ri )\nn\\&&\lf .
+\ln 'c_T\lf \{\f{\p ''}{\p '}\ln '\lf (\f{c_Tc_X}{c_R}\ri )+\ln''\lf (\f{c_Tc_X}{c_R}\ri )-\f{2c_R^2\cP '}{\b\p '}\ln 'c_R-\f{c_R^2\cP''}{\b}
\ri \}
\ri ].
\eea
Using equations (\ref{Dn1}) and (\ref{En2}), we can write
\bea
\f{c_R^2\cP '}{\b}\ln 'c_T\ln 'c_R&=&2\p ''\f{c_T'}{c_T}\f{c_R'}{c_R}
\nn\\&& +
2\ln '\lf (\f{c_Tc_X}{c_R}\ri )\lf \{\ln ''c_T +\lf (\f{c_T'}{c_T}\ri )^2+\f{\b}{2}\p ^{\pr 2}+\f{c_R^2\cP}{2}\ri \}.
\eea
Using it, we get
\bea
K'&=&-\f{2\p ''}{\p '}K-2c_R^2\cP '-\f{2c_R^2\cP ''}{\b\p '}\ln 'c_T\nn\\&&
+\f{2}{\p '}\lf [\ln 'c_T\lf \{\f{\p ''}{\p '}\ln '\lf (\f{c_T^3}{c_Xc_R^3}\ri )
-\ln ''c_X-2\ln 'c_T\ln '\lf (\f{c_Tc_X}{c_R}\ri )
\ri \}\ri .\nn\\&& \lf .
+\ln ''c_T\ln '\lf (\f{c_Tc_R^3}{c_X^4}\ri )-c_R^2\cP \ln '(c_Tc_X)
\ri ].
\eea
The left hand side of the equation (\ref{Atphi2}) then becomes
\bea
LHS&=&\f{2}{\p '}\lf [\ln ''c_T\ln '\lf (\f{c_T^4}{c_X^5}\ri )+2\ln '\lf (\f{c_T}{c_X}\ri )\ln 'c_T\ln '\lf (\f{c_T^3}{c_Xc_R^3}\ri )+2\b\p ^{\pr 2}\ln '\lf (\f{c_T}{c_X}\ri )\ri .\nn\\&&\lf .-c_R^2\cP\ln '(c_Tc_X)-\ln 'c_T\ln ''c_X -2(\ln 'c_T)^2\ln '\lf (\f{c_Tc_X}{c_R}\ri )\ri ].
\eea
Using equation (\ref{ctcxpr}), we get
\bea
LHS&=&\f{2}{\p '}\lf [\ln '\lf (\f{c_T}{c_X}\ri )\lf \{4\ln ''c_T +2\ln 'c_T\ln '\lf (\f{c_T^3}{c_Xc_R^3} \ri )+2\b\p ^{\pr 2}\ri \}\ri .\nn\\&&
\lf .+\ln '\lf (\f{c_Tc_X}{c_R}\ri )\lf \{-c_R^2\cP +\f{\b\p ^{\pr 2}}{2}-2(\ln 'c_T)^2\ri \}\ri ].\nn\\
\eea
Using equation (\ref{lnct2}), we get
\bea
LHS&=&\f{2}{\p '}\lf [\ln '\lf (\f{c_T}{c_X}\ri )\lf \{\b\p ^{\pr 2}-2c_R^2\cP +2\ln 'c_T\ln '\lf (\f{c_T}{c_Xc_R} \ri )\ri \}\ri .\nn\\&&
\lf .+\ln '\lf (\f{c_Tc_X}{c_R}\ri )\lf \{-c_R^2\cP +\f{\b\p ^{\pr 2}}{2}-2(\ln 'c_T)^2\ri \}\ri ]\nn\\
&=&\f{2}{\p '}\lf [\ln '\lf (\f{c_T}{c_X}\ri )\lf \{4\ln 'c_T\ln 'c_X +2\ln 'c_T\ln '\lf (\f{c_T}{c_Xc_R} \ri )\ri \}\ri .\nn\\&&
\lf .+\ln '\lf (\f{c_Tc_X}{c_R}\ri )\lf \{2\ln 'c_T\ln 'c_X -2(\ln 'c_T)^2\ri \}\ri ]\nn\\
&=&0=RHS,
\eea
where we used equation (\ref{En3}) to get middle step.

\bibliography{d1-hydro}
\bibliographystyle{JHEP}

\end{document}